\documentclass[twocolumn,apj]{emulateapj}
\usepackage{apjfonts}

\shorttitle{Decomposing Star Formation and AGN} 
\shortauthors{Fu et al.}
\journalinfo{to appear in the Astrophys. J.}
\submitted{Received 2010 June 15; Accepted 2010 August 17}

\usepackage[pagebackref=false,letterpaper=true,colorlinks=true,citecolor=black,linkcolor=blue,breaklinks=true,bookmarks=true]{hyperref}

\newcommand{\eg}{e.g.,}
\newcommand{\ie}{i.e.,}
\newcommand{\spitzer}{{\it Spitzer}}
\newcommand{\chandra}{{\it Chandra}}
\newcommand{\hst}{{\it HST}}
\newcommand{\um}{$\mu$m}
\newcommand{\uJy}{$\mu$Jy}
\newcommand{\vmax}{$V_{\rm max}$}
\newcommand{\ergs}{{erg s$^{-1}$}}
\newcommand{\msun}{$M_{\odot}$}
\newcommand{\lsun}{$L_{\odot}$}
\newcommand{\zbin}{$0.6 < z < 0.8$}
\newcommand{\kcors}{{\it K}-corrections}
\newcommand{\kcor}{{\it K}-correction}
\newcommand{\Lfif}{$L_{15}$}
\newcommand{\Leig}{$L_{8}$}
\newcommand{\Lir}{$L_{\rm IR}$}

\newcommand{\iab}{$i_{\rm AB}^{\,+}$}
\newcommand{\pagn}{$p_{\rm AGN}$}

\begin{document}

\title{Decomposing Star Formation and Active Galactic Nucleus with {\em Spitzer} Mid-Infrared Spectra: Luminosity Functions and Co-Evolution} 
\author{
Hai Fu\altaffilmark{1}, 
Lin Yan\altaffilmark{2},
N.\ Z.\ Scoville\altaffilmark{1},
P.\ Capak\altaffilmark{3}, 
H.\ Aussel\altaffilmark{4}, 
E.\ Le Floc'h\altaffilmark{4},
O.\ Ilbert\altaffilmark{5},
M.\ Salvato\altaffilmark{6},
J.\ S.\ Kartaltepe\altaffilmark{7},
D.\ T.\ Frayer\altaffilmark{8}, 
D.\ B.\ Sanders\altaffilmark{9},
K.\ Sheth\altaffilmark{10},
and Y.\ Taniguchi\altaffilmark{11}
}

\altaffiltext{1}{Astronomy Department, California Institute of Technology, MS 249$-$17, Pasadena, CA 91125, USA} 
\altaffiltext{2}{{\em Spitzer} Science Center, California Institute of Technology, MS 220$-$06, Pasadena, CA 91125, USA}  
\altaffiltext{3}{Infrared Processing and Analysis Center, California Institute of Technology, MS 100$-$22, Pasadena, CA 91125, USA}
\altaffiltext{4}{Laboratoire AIM-Paris-Saclay, CEA/DSM/Irfu - CNRS - Universit\'e Paris Diderot, CE-Saclay, pt courrier 131, F-91191 Gif-sur-Yvette, France} 
\altaffiltext{5}{Laboratoire d'Astrophysique de Marseille, BP 8, Traverse du Siphon, 13376 Marseille Cedex 12, France}
\altaffiltext{6}{IPP - Max-Planck-Institute for Plasma Physics, Boltzmannstrasse 2, D-85748, Garching, Germany}
\altaffiltext{7}{National Optical Astronomy Observatory, 950 North Cherry Avenue, Tucson, AZ 85 721, USA} 
\altaffiltext{8}{National Radio Astronomy Observatory, PO Box 2, Green Bank, WV 24944, USA}
\altaffiltext{9}{Institute for Astronomy, University of Hawaii, Honolulu, HI 96822, USA}
\altaffiltext{10}{Department of Astronomy, University of Maryland, College Park, MD 20742, USA}
\altaffiltext{11}{Research Center for Space and Cosmic Evolution, Ehime University, 2-5 Bunkyo-cho, Matsuyama 790-8577, Japan}

\begin{abstract} 
We present {\it Spitzer} 7$-$38\,\um\ spectra for a 24\,\um\ flux limited sample of galaxies at $z \sim 0.7$ in the COSMOS field. The detailed high-quality spectra allow us to cleanly separate star formation (SF) and active galactic nucleus (AGN) in individual galaxies. We first decompose mid-infrared Luminosity Functions (LFs). We find that the SF 8\,\um\ and 15\,\um\ LFs are well described by Schechter functions. AGNs dominate the space density at high luminosities, which leads to the shallow bright-end slope of the overall mid-infrared LFs. The total infrared (8$-$1000\,\um) LF from 70\,\um\ selected galaxies shows a shallower bright-end slope than the bolometrically corrected SF 15\,\um\ LF, owing to the intrinsic dispersion in the mid-to-far-infrared spectral energy distributions. We then study the contemporary growth of galaxies and their supermassive black holes (BHs). Seven of the 31 Luminous Infrared Galaxies with {\it Spitzer} spectra host luminous AGNs, implying an AGN duty cycle of $23\pm9$\%. The time-averaged ratio of BH accretion rate and SF rate matches the local $M_{\rm BH}$$-$$M_{\rm bulge}$ relation and the $M_{\rm BH}$$-$$M_{\rm host}$ relation at $z \sim 1$. These results favor co-evolution scenarios in which BH growth and intense SF happen in the same event but the former spans a shorter lifetime than the latter. Finally, we compare our mid-infrared spectroscopic selection with other AGN identification methods and discuss candidate Compton-thick AGNs in the sample. While only half of the mid-infrared spectroscopically selected AGNs are detected in X-ray, $\sim$90\% of them can be identified with their near-infrared spectral indices.
\end{abstract}


\section{Introduction}\label{intro}

A central question in extragalactic research is the mass assembly history of galaxies and their supermassive black holes (BHs). Understanding this process requires measurements of the rates of star formation (SF) and BH accretion throughout the cosmic ages. As both processes typically occur in dusty environments, the majority of the energy emerges as dust-reprocessed thermal infrared (IR) emission. The IR luminosity function (LF), defined as the density distribution over IR luminosity, thus holds clues to the evolution of galaxies and BHs. 

Measuring IR LFs has a long history. About two decades ago, surveys with the {\it Infrared Astronomical Satellite} ({\it IRAS}) established the local benchmarks in the mid-IR and far-IR wavelengths \citep[$\lambda > 5$\,\um; \eg][]{Rush93, Soifer87}. Although the space density of Luminous and Ultra-Luminous Infrared Galaxies (LIRGs/ULIRGs; log(\Lir/\lsun) = log[$L$(8$-$1000\um)/\lsun] = [11, 12]/[12, $\infty$)) exceeds those of optically-selected galaxies and QSOs at comparable bolometric luminosities, they account for only $\sim$5\% of the total integrated IR energy density in the local universe \citep{Soifer91}. LIRGs, however, dominate SF activities at higher redshifts. Strong evolution of IR-selected population with look-back time was suggested by the number count results from the {\it Infrared Space Observatory} ({\it ISO}) observations \citep[][]{Elbaz99}. Remarkably, this evolution was first detected in the limited redshift range covered by the earlier {\it IRAS} surveys \citep[$0 < z < 0.2$; \eg][]{Saunders90}. Recently, many attempts have been made to estimate IR LFs at redshifts up to $z \sim 3$ using data from the new generation of IR space telescopes---\spitzer\ \citep{Le-Floch05,Perez-Gonzalez05,Franceschini06,Bell07,Caputi07,Magnelli09,Rodighiero10}, {\it AKARI} \citep{Goto10}, and {\it Herschel} \citep{Gruppioni10,Eales10}. The strong evolution was confirmed; \citet{Le-Floch05} concluded that the 24\,\um-derived IR (8$-$1000\,\um) comoving energy density evolves as $(1+z)^{4}$ at $0 < z < 1$ and LIRGs contribute $\sim$70\% of the energy density at $z = 1$.

Although a crude picture has already emerged, our knowledge of the galaxy LFs is still limited by two main factors: (1) the poorly sampled IR spectral energy distribution (SED), and (2) the elusive contribution from active galactic nuclei (AGNs). Constrained by the low sensitivities of existing far-IR and sub-milimeter instruments, previously published 8$-$1000\,\um\ LFs at $z \gtrsim 0.3$ are mostly {\it K}-corrected mid-IR LFs. The \kcor\ depends strongly on the assumed SED, which without long wavelength data one can only guess from the SEDs of local galaxies. By opening the far-IR window, {\it Herschel} will soon revolutionize this field. On the other hand, besides dust-enshrouded SF, mid-IR emission can also be powered by AGNs, as their accretion disks can heat the surrounding dusty tori. One therefore should separate AGN and SF before using mid-IR data to build 8$-$1000\,\um\ LFs. X-ray imaging has been widely used to identify AGNs but the identification becomes incomplete at moderately high redshifts, because of the insufficient depth in hard X-ray, which has less biases against absorbed AGNs. In addition, the significant population of AGN/SF composite systems renders it problematic to simply remove all of the identified AGNs from the sample. Only with mid-IR spectroscopy we can decompose AGN and SF in the mid-IR LFs by separating the two in individual sources.

AGNs exhibit a class of IR SEDs distinct from star-forming galaxies \citep{Hao07,Richards06,Netzer07,Fu09b}, due to the clumpy nature of the torus \citep{Sirocky08,Nikutta09}. Broadly speaking, AGN SEDs ($\nu L\nu$) are flat from near-IR to mid-IR but drop beyond 20\,\um, and SF SEDs peak between 55$-$120\,\um\ \citep{Rieke09}.

The shape of the LF depends on the rest-frame wavelength. From UV to near-IR wavelengths, the classical \citet{Schechter76} function provides excellent fits \citep[\eg][]{Arnouts05,Ilbert05,Babbedge06,Dai09}:

\begin{equation}
\phi(L)=\frac{dN(L)}{\;dV\;d{\log}(L)}=\phi^{\star} \left(\frac{L}{L^{\star}}\right)^{1+\alpha}\exp\left(-\frac{L}{L^{\star}}\right)
\end{equation}

\noindent 
At mid-IR and far-IR wavelengths, the observed bright-end slope ($\phi \propto L^{-2}$) is shallower than the exponential cut-off of the Schechter function \citep[\eg][]{Rush93,Lawrence86}. These LFs are best described by {\it modified Schechter functions} \citep[][]{Saunders90}: 

\begin{equation}
\phi(L)=\phi^{\star} \left(\frac{L}{L^{\star}}\right)^{1+\alpha}\exp\left[-\frac{1}{2{\sigma}^2}{\log}^2\left(1+\frac{L}{L^{\star}}\right)\right]\nonumber
\end{equation}

It may appear counter-intuitive that UV LFs show a profile different from that of far-IR LFs, since both UV and far-IR emission trace active SF. One explanation is that since the UV LF is not extinction corrected, its bright end is steepened because dust extinction increases with star formation rate \citep[SFR; ][]{Hopkins01,Martin05,Zheng07a,Reddy10}. 

AGNs might be responsible for the shallow bright-end slope of the mid-IR LFs. \citet{Rush93} showed that the local 12\,\um\ LF of normal galaxies (\ie\ excluding Seyferts of both spectral types) has a steeper bright-end slope ($\phi \propto L^{-3.6}$). Additionally, after removing spectroscopically classified AGNs, \citet{Huang07} found the local 8\,\um\ LF was well fit by a Schechter function \citep[see also][]{Dai09}. At higher redshifts, the results are less conclusive, because of the limitations from both the small survey areas and the uncertainties in AGN identification. 

In this paper, we decompose AGN and SF in mid-IR LFs at $z \sim 0.7$ using a unique spectroscopic data-set of LIRGs. We extend these results to lower luminosities (\Lir\ $> 10^{10}$ \lsun) with the 2 deg$^2$ S-COSMOS 24\,\um\ catalog \citep{Sanders07,Le-Floch09}. In \S~\ref{data} we present \spitzer\ Infrared Spectrograph \citep[IRS;][]{Houck04} spectra of a 24\,\um\ flux limited sample at $z \sim 0.7$ and cross-match COSMOS multi-wavelength photometry and redshift catalogs. In \S~\ref{irs}, we model the IRS spectra to divide the observed luminosities between AGN and SF. This method, for the first time, allows us to consider all of the sources in the sample and derive AGN and SF LFs separately (\S~\ref{LF}). To estimate the total SFR function, we use a 70\,\um\ selected sample to derive the 8$-$1000\,\um\ LF and compare it with the SF mid-IR LF (\S~\ref{IRLF}). In \S~\ref{discussion}, we explore galaxy$-$BH co-evolution in LIRGs, compare our mid-IR spectroscopic AGN identification with photometric AGN selections, and discuss the possible presence of Compton-thick AGNs in our sample. We close by summarizing our conclusions in \S~\ref{conclusion}.

Throughout we assume a cosmological model with $H_0=70$ km s$^{-1}$ Mpc$^{-1}$, $\Omega_m=0.3$, and $\Omega_{\Lambda}= 0.7$. All magnitudes are in the AB system. The SFRs are given for the \citet{Chabrier03} initial mass function (IMF); they can be converted to the \citet{Salpeter55} IMF by adding $\sim$0.24 dex to the logarithm.

\begin{figure*}[!t]
\epsscale{1.15}
\plotone{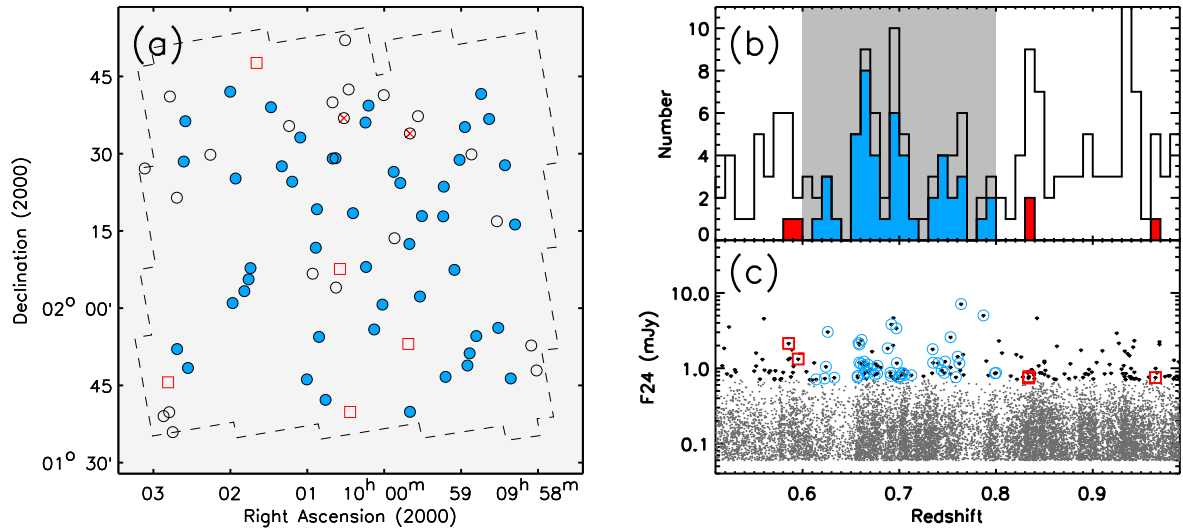}
\caption{IRS sample selection and completeness. (a) Distribution of the IRS sample in the COSMOS field. The dashed polygon delineates the \hst/ACS coverage. Black open circles mark the 70 MIPS sources at \zbin\ and $F24 > 0.7$ mJy inside the ACS field. 48 of those have IRS spectra and are filled in blue. Red open squares mark the 5 IRS sources that fall outside of the redshift window. The red crosses show the 2 IRS sources for which the observations failed due to peak-up acquisition errors. (b) Redshift distribution of the MIPS 24\,\um\ sample with $F24 > 0.7$ mJy and inside the ACS field (black solid histogram), in comparison with that of the IRS sample (blue/red filled histogram). Grey shaded region shows our redshift selection window. (c) 24\,\um\ flux vs. redshift. MIPS sources inside the ACS field with $F24 > 0.7$ mJy and $F24 < 0.7$ mJy are shown as black and grey points, respectively. The blue open circles and red open squares highlight the 53 sources with IRS spectra. 
\label{fig:sample}} 
\end{figure*}

\section{Data}\label{data}

The data required to build mid-IR LFs include: (1) IRS spectroscopy of a complete 24\,\um\ flux limited sample ($F24 > 0.7$\,mJy), (2) the Multiband Imaging Photometer for \spitzer\ \citep[MIPS;][]{Rieke04} 24\,\um\ catalog down to 60\,\uJy\ \citep{Sanders07,Le-Floch09}, and (3) the redshift catalogs (both spectroscopic and photometric redshifts). The IRS spectra were used to decompose AGN and SF. The MIPS catalog was used to build LFs down to the luminosity limit of the S-COSMOS survey. In order to remove stellar photospheric emission in the mid-IR, we also included the 30-band COSMOS photometry from Galex/$FUV$ (1551\AA) to Spitzer/IRAC 8\,\um. For the 8$-$1000\,\um\ LF, we used the MIPS 70/160\,\um\ data \citep{Frayer09}. In this section, we describe our IRS observations and explain the multi-wavelength cross-matching method.

\subsection{IRS Spectroscopy}\label{irsobs}

We selected the targets of our Cycle 5 \spitzer/IRS Legacy program (PID 50286, PI: Scoville) using the following criteria: (1) they are inside the COSMOS field observed by {\it Hubble Space Telescope}/Advanced Camera for Surveys (\hst/ACS), (2) have MIPS 24\,\um\ flux brighter than 0.7 mJy, and (3) have photometric redshifts between $0.6 < z < 0.8$. Nine blended sources were removed to avoid confusion in the spectra. AORs were generated for 55 galaxies. 53 were successfully observed, and 2 targets failed due to errors in the peak-up imaging acquisition (Table~\ref{tab:sample}). 

At the time of writing, 48 of the 53 galaxies (90\%) have spectroscopic redshifts (including eight new redshifts from the IRS spectra). In Figure~\ref{fig:sample} we show the distributions of this sample in the COSMOS field, in redshift, and in 24\,\um\ flux. We compare these distributions with the parent sample using the currently best-estimated redshifts (\S~\ref{dataset}). We have spectroscopic redshifts for 49\% of the 24\,\um\ sources above 0.7\,mJy and for 79\% of the X-ray detected 24\,\um\ sources above 0.7\,mJy. Most of the spectra come from zCOSMOS \citep{Lilly07} and the COSMOS AGN Spectroscopic Survey \citep{Trump09}. We found 70 sources above 0.7\,mJy between \zbin\ over the 1.63 deg$^2$ \hst/ACS field \citep{Scoville07b}. 48 of the 70 (68.6\%) have IRS spectra. Accounting for this difference, the effective area of the IRS sample is 1.12 deg$^2$. We included the 5 IRS sources outside of \zbin\ in the spectral analysis (\S~\ref{irs}), although we ignored them when estimating LFs.  

Low-resolution IRS spectra ($R\sim60-130$) were taken in January 2009. We used the Short Low module in 1st order (SL1, 7.4$-$14.5\,\um) and the Long Low module in 1st (LL1, 19.5$-$38\,\um) and 2nd order (LL2, 14.0$-$21.3\,\um) to continuously cover a wavelength range of 7.4$-$38\,\um. All observations were made in the standard staring mode. One cycle consisted of two nod positions offset by 20\arcsec\ and 55\arcsec\ for SL and LL modules, respectively. The number of cycles was adjusted to yield approximately equal signal-to-noise ratios across modules and sources. Between 2 and 25 cycles of 60\,s, 2 and 15 cycles of 120\,s, 2 and 10 cycles of 120\,s were used in SL1, LL2, and LL1, respectively (Table~\ref{tab:sample}). 

The data were first passed through the S18.7 version of the SSC pipeline. Spectra were then extracted from the BCD files using the IRSLow IDL pipeline \citep{Fadda10}. Briefly, the pipeline procedure is as follows: (1) After rejecting bad pixels and subtracting background from individual frames, the pipeline combines all of the 2D spectra at different nod positions; (2) Optimal background subtraction is done in the second pass by manually masking off objects from the background-subtracted coadded image; (3) 1D spectra are extracted using PSF (Point Spread Function) fitting with PSF profiles from IRS calibration stars. Lastly, we removed bad portions near the edges of each order from the 1D spectra.

The LL spectra of MIPS\,31635 and 46960 appear unusable. The former is contaminated by a nearby unidentified bright object (likely an asteroid), and the latter is dominated by a featureless continuum $\sim5\times$ higher than the MIPS 24\,\um\ flux. We discarded these spectra and used their MIPS 24\,\um\ photometry as surrogates.

In order to check the flux calibration, we measured 24\,\um\ fluxes from the IRS spectra by convolving them with the MIPS filter transmission curve. Note that at $z \sim 0.7$, our objects are unresolved in the 24\,\um\ image, and the flux loss due to the finite slit widths\footnote{Slit widths are 3\farcs7, 10\farcs5, and 10\farcs7 for SL1, LL2, and LL1 modules, respectively.} have been fully accounted for during the spectral extraction. We found $F24_{\rm IRS}/F24_{\rm MIPS} = 1.078\pm0.096$. The MIPS fluxes are systematically lower than the IRS fluxes mainly because they are calibrated against a 10,000 K blackbody instead of a flat $F_{\nu}$ as in the AB system. As we determined our \kcors\ for the 24\,\um\ sources using the IRS spectra (\S~\ref{completeness}), we multiplied the MIPS 24\,\um\ fluxes by 1.078 for sources between \zbin.

\subsection{Multi-Wavelength Dataset}\label{dataset}

\begin{figure*}[!t]
\epsscale{1.15}
\plotone{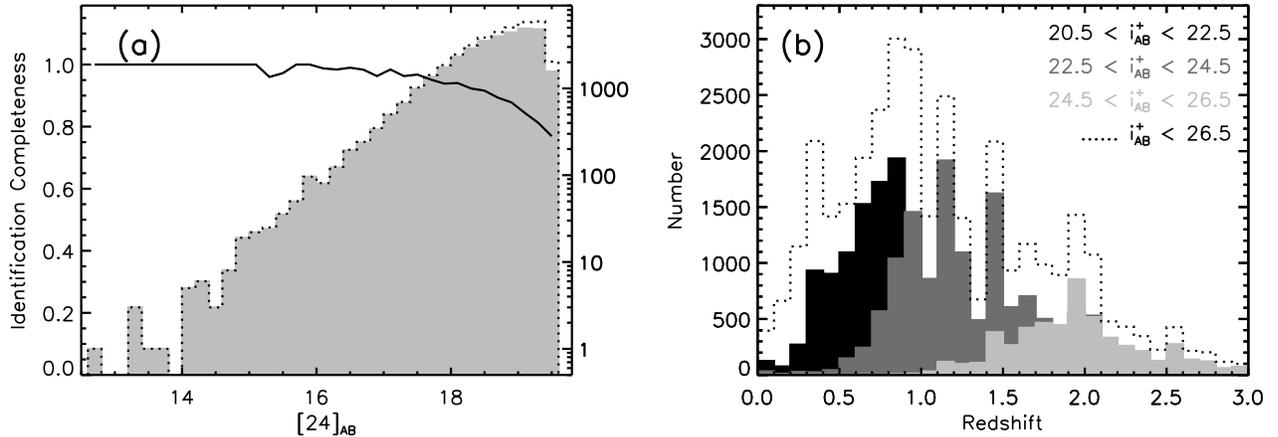}
\caption{The optical identification of 24\,\um\ sources above 60\,\uJy\ and between \zbin\ is $\sim$99\% complete. (a) Identification completeness of 24\,\um\ sources at {\it all} redshifts as a function of 24\,\um\ AB magnitudes ({\it solid line}). Dotted histogram shows the distribution of the 39,441 objects within the Subaru deep area but outside of the masked areas, and the light grey filled histogram shows the distribution of the 34,883 objects with secure optical counterparts. The identification completeness is simply the ratio of the two. (b) Redshift distributions of 24\,\um\ sources with optical counterparts at $20.5 <$\,\iab\,$< 22.5$ ({\it black}),  $22.5 <$\,\iab\,$< 24.5$ ({\it grey}), $24.5 <$\,\iab\,$< 26.5$ ({\it light grey}),  and \iab\,$< 26.5$ ({\it dotted histogram}). 
\label{fig:completeness}} 
\end{figure*}

Redshift is critical for estimating LFs. We matched the MIPS 24\,\um\ catalog \citep{Le-Floch09} with the COSMOS dataset to identify redshifts. The full MIPS catalog contains 52,092 sources down to 60\,\uJy. The cross-matching was performed inside the 2 deg$^2$ Subaru deep area ($149.4114075^{\circ} < \alpha < 150.8269348^{\circ}$ and $1.4987870^{\circ} < \delta < 2.9127350^{\circ}$). The optical counterparts of 24\,\um\ sources were identified in two steps. First, the 24\,\um\ coordinates were matched to the nearest IRAC detection within a 2\arcsec\ radius \citep{Ilbert10}. Then, we searched for the nearest optical counterpart at \iab\,$< 26.5$ within 1\arcsec\ to the IRAC position. We used the combined $FUV$-to-$K_S$ photometric catalog \citep{Capak07} so that the $FUV$-to-24\,\um\ SEDs were assembled at the same time. When spectroscopic redshifts were unavailable\footnote{15\% of the 24\,\um\ sources in the 60\,\uJy\ catalog have spectroscopic redshifts.}, we used the photometric redshifts of \citet{Ilbert09} and \citet{Salvato09} for normal galaxies and X-ray sources, respectively. Lastly, we removed sources inside the Subaru/optical and \spitzer/IRAC 3.6\,\um\ masks, where the quality of photometry and photometric redshifts degrades due to the proximity to very bright objects. The area of this final catalog is 1.66 deg$^2$ (the masked area is 0.28 deg$^2$), enclosing a total of 39,525 MIPS sources down to 60\,\uJy. Among these, 34,967 have secure optical counterparts.

We found 84 instances of source fragmentation in the MIPS catalog, and we combined the MIPS fluxes for these sources. The MIPS-selected sample was thus reduced to a total of 39,441 objects with 34,883 secure counterparts (88.4\%). Figure~\ref{fig:completeness}$a$ shows the optical identification completeness as a function of 24\,\um\ magnitudes. Note that the completeness reaches 75\% even for objects as faint as [24]$_{\rm AB}$ = 19.45 (= 60\,\uJy).

To build the integrated 8$-$1000\,\um\ LF, we also identified closest MIPS 70/160\,\um\ counterparts of 24\,\um\ sources using the MIPS-Germanium 3$\sigma$ (peak-to-noise) catalog \citep{Frayer09}. We adopted a matching radius that equals the cataloged 2$\sigma$ radial positional error from PSF fitting. Following \citet{Kartaltepe10}, 559 spurious 70\,\um\ sources were removed from the catalog. We found 70/160\,\um\ counterparts for 1831/528 of the 34,883 24\,\um\ sources that are optically identified. Our counterpart identification covers a $\sim$50\% larger area than that of \citet{Kartaltepe10}, since we did not restrict it to the ACS field; but both 70\,\um\ samples produce consistent 8$-$1000\,\um\ LFs (\S~\ref{IRLF}).

\section{Decomposition of IRS Spectra}\label{irs}

\begin{figure*}[!b]
\epsscale{1.2}
\plotone{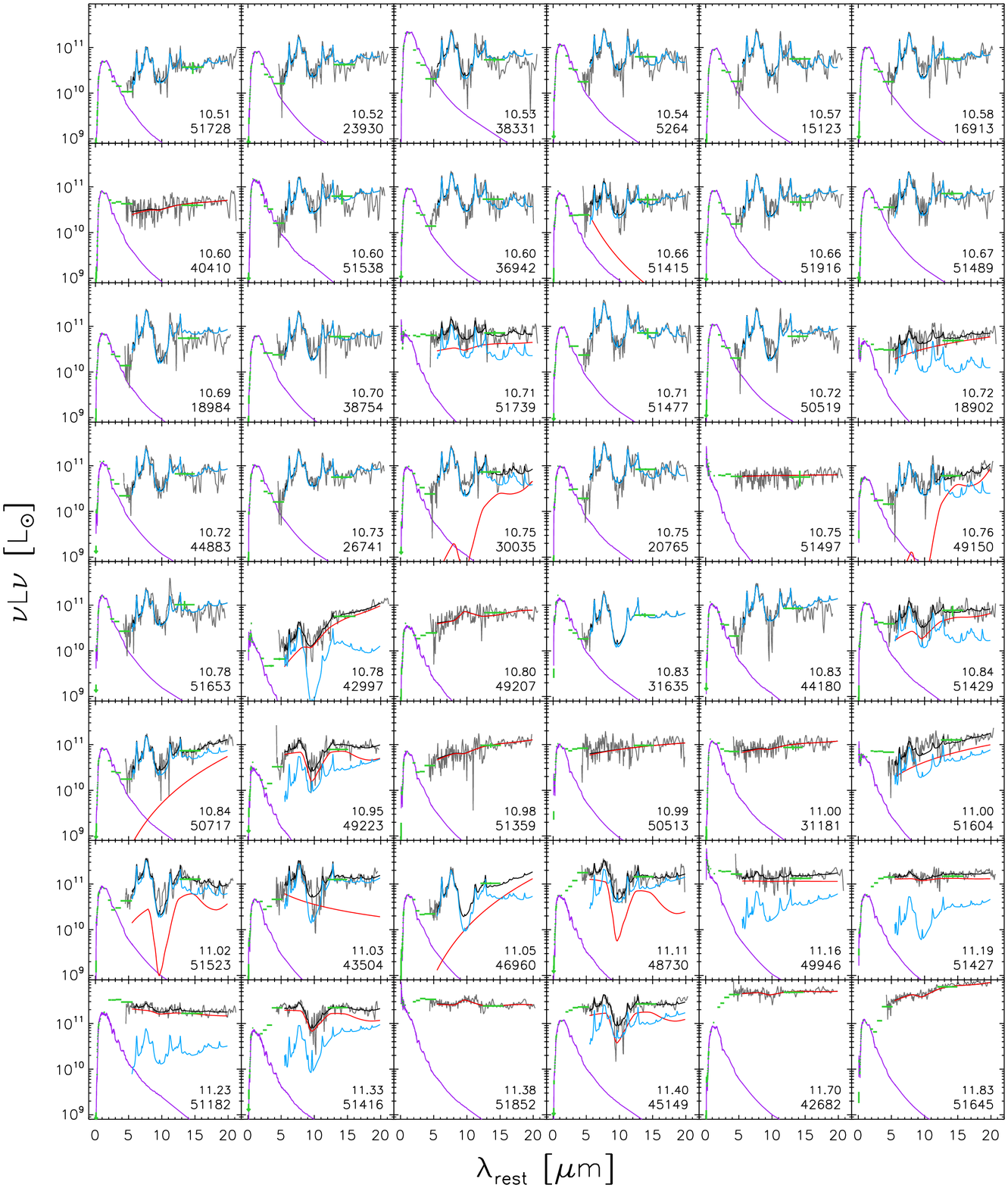}
\caption{Decomposition of AGN and SF with IRS spectra. Panels are sorted by ascending rest-frame 15\,\um\ luminosity to show the dramatic increase of AGN contribution with luminosity. Labeled are log(\Lfif/\lsun) and MIPS IDs. IRS spectra are shown in grey and photometric data are green horizontal bars. The black curve shows our best-fit model to the IRS spectra, which is a combination of stellar photospheric emission ({\it purple}), dust-reprocessed SF ({\it blue}) and power-law AGN ({\it red}). 
\label{fig:irs}} 
\end{figure*}

\begin{figure}[!tb]
\epsscale{1.2}
\plotone{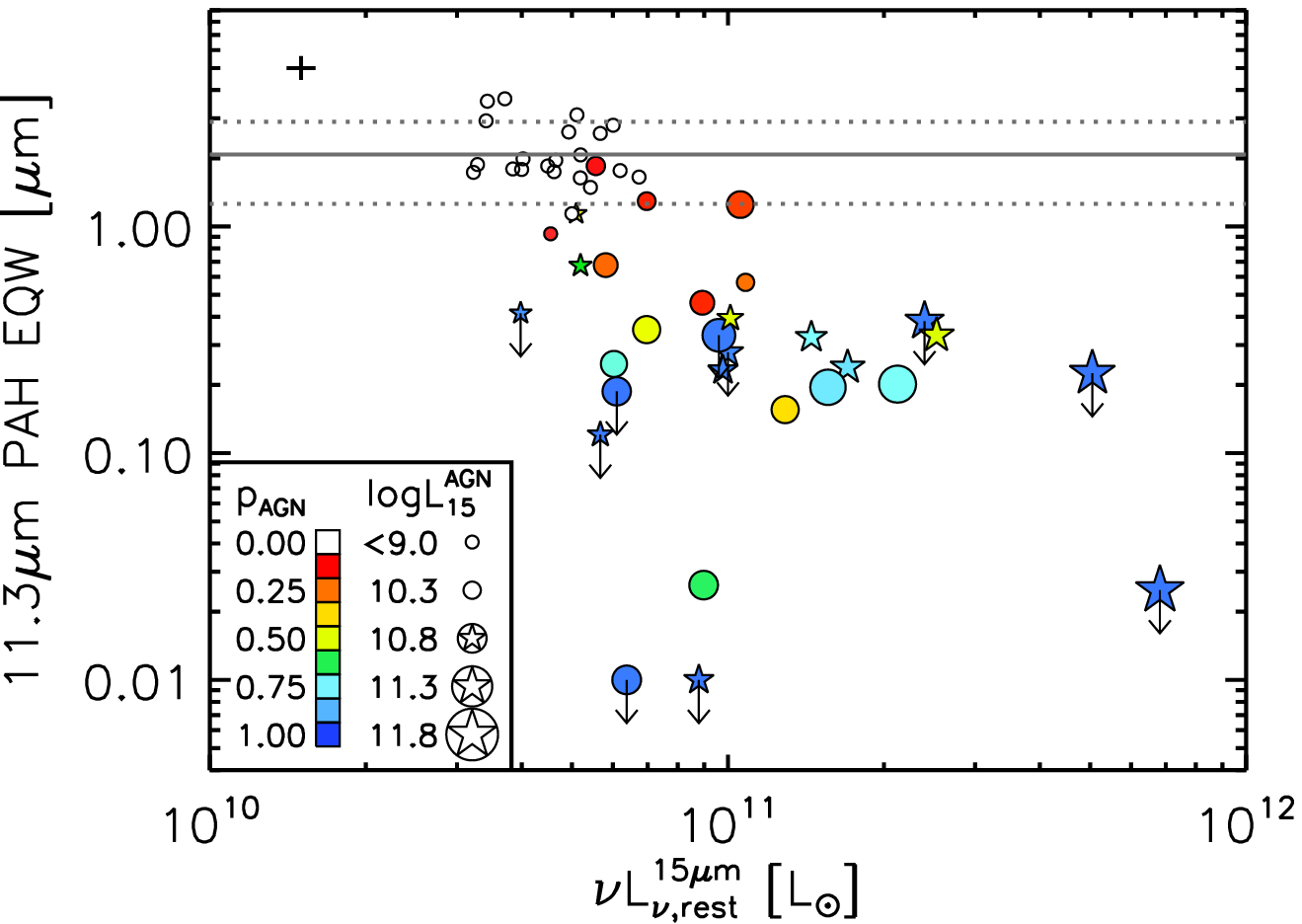}
\caption{PAHFit 11.3\,\um\ PAH EQW vs. rest-frame 15\,\um\ luminosity (\Lfif). The IRS-selected AGNs ({\it color-filled symbols}) have lower PAH EQWs and higher luminosity, similar to local warm ULIRGs \citep[][]{Genzel98,Armus07}. X-ray detected sources are plotted as stars. The colors of the symbols are coded according to the AGN contribution integrated over $5.5 < \lambda_{\rm rest} < 20$\,\um\ (\pagn), and the sizes increase logarithmically with the AGN luminosity at rest-frame 15\,\um\ ($L_{15}^{\rm AGN}$). The median errors are shown at the top left corner. The upper limits indicate spurious PAH detections. The solid and dotted lines show the mean and 1$\sigma$ dispersion of the EQWs of pure star-forming galaxies (\pagn\ = 0). 
\label{fig:paheqw}}
\end{figure}

Mid-IR spectra can be decomposed into SF and AGN components \citep[\eg][]{Sajina07,Pope08}. Photospheric emission from low-mass stars can be significant below 10\,\um. To estimate its contribution in the IRS spectra, we fit the rest-frame 912 \AA\ to 1.3\,\um\ SED (\ie\ GALEX $FUV$ to $K_S$ bands) using \citet{Bruzual03} stellar population synthesis models and extrapolated the best-fit SED to the mid-IR. Once the photospheric emission was subtracted, we modeled the IRS spectrum as a superposition of dust-obscured SF and AGN. The former is dominated by polycyclic aromatic hydrocarbons (PAH) features, and the latter can be approximated by a power law. For the SF component, we used the averaged templates of \citet{Rieke09}, built from nearby star-forming galaxies in 14 luminosity bins, 9.75 $<$ log(\Lir/\lsun) $<$ 13. Both the AGN and SF components were subjected to attenuation from a screen of dust, adopting the modified Galactic center extinction curve of \citet{JDSmith07} with $\beta = 0.1$. 

Our fitting procedure is as follows. First, we corrected the templates for silicate extinction and stellar emission with PAHFit \citep{JDSmith07}. We used the same extinction curve and approximated the stellar emission with a 5,000 K blackbody. Then, we found the best solution for each of the 14 templates using MPFIT, an IDL $\chi^2$-minimization routine \citep{Markwardt09}. As we allowed different amount of extinctions for the two components, there are five free parameters in each model --- power-law index, scaling factors for the two components, and their extinctions at 9.7\,\um\ ($\tau_{9.7}^{\rm AGN}$, $\tau_{9.7}^{\rm SF}$). Finally, the solution giving the global minimum $\chi^2$ was selected as the best two-component model.

Since many sources appear dominated by either SF or AGN, we also performed least-$\chi^2$ fits with a single component. In the AGN-only model we fit for both silicate emission and absorption. The $\chi^2$ values from the two single-component models were compared with that of the two-component model, and the best model was determined using an \textsl{F} test. The two-component model was deemed necessary only when its $\chi^2$ reduction relative to the single-component models yielded an \textsl{F} value that is less than 5\% likely (\ie\ the probability that the $\chi^2$ reduction is due to random error is less than 5\%). We used SF-only, AGN-only, and composite models in 22, 10, and 21 sources, respectively. 

We sorted our modeling results by ascending rest-frame 15\,\um\ luminosity (\Lfif) for the 48 sources between \zbin\ in Figure~\ref{fig:irs}. The AGN contribution increases dramatically with luminosity: while they are absent in less luminous sources (log(\Lfif) $<$ 10.6 \lsun\ or $F24 \lesssim 0.8$\,mJy), AGNs dominate above log(\Lfif) $>$ 10.9 \lsun ($F24 \gtrsim 1.2$\,mJy). The sources hosting AGNs show IRAC fluxes clearly above the best-fit stellar population synthesis models; and the amount of the excess agrees well with the best-fit power-law model. At longer wavelengths, since a typical AGN SED declines sharply beyond rest-frame 20\,\um, one would expect the far-IR emission to be dominated by SF. For the 15 MIPS 70\,\um-detected SF-only sources, although the extrapolation from the best-fit SF template generally over-predicts the 70\,\um\ flux, the difference is within 0.5 dex. The disagreement is more severe for the SF/AGN composite systems, and the three 70\,\um-detected AGN-only sources all have 70\,\um\ fluxes in excess of those extrapolated from the mean AGN SED of \citet{Netzer07}, hinting that AGNs could also be a significant power source of the rest-frame 40\,\um\ emission. We will discuss further on the far-IR emission in \S~\ref{IRLF}.

As a measure of the strength of SF relative to that of AGN heated dusts, PAH equivalent width (EQW) is recognized as a good AGN diagnostic \citep{Genzel98,Armus07}. In Fig.~\ref{fig:paheqw} we show that the 11.3\,\um\ PAH EQW decreases as AGN intensifies, confirming that our spectral decomposition and the PAH EQWs are equivalent in identifying AGNs. Note that the PAH EQWs were measured with PAHFit \citep{JDSmith07}, which yields EQWs a few times larger than the spline method \citep{Fadda10}. We will present comparisons with other commonly used AGN diagnostics in \S~\ref{AGN}.

\section{Luminosity Functions} \label{LF}

\subsection{Completeness, \kcor\ and 1/\vmax\ Estimator} \label{completeness}

\begin{figure*}[!t]
\epsscale{0.58}
\plotone{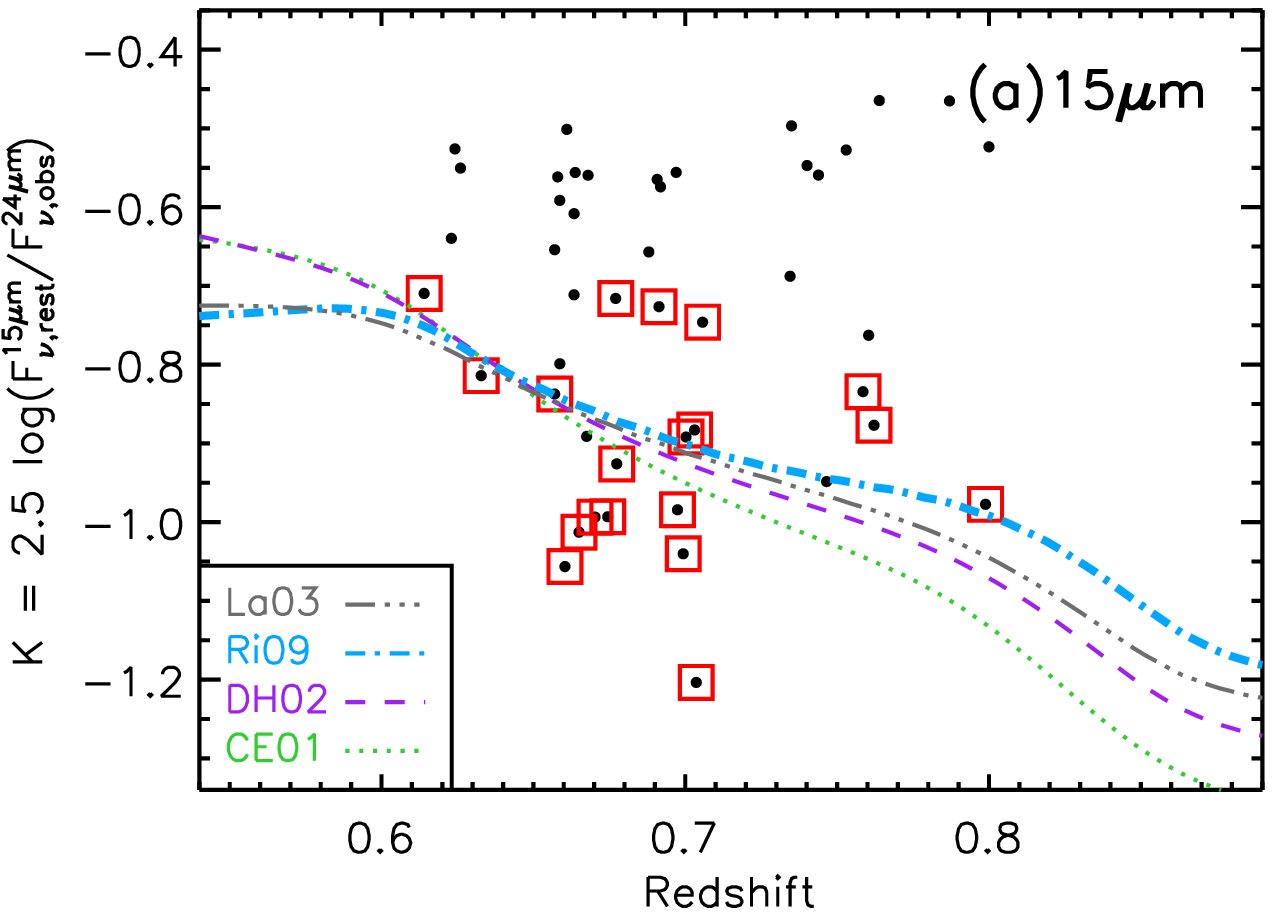}
\plotone{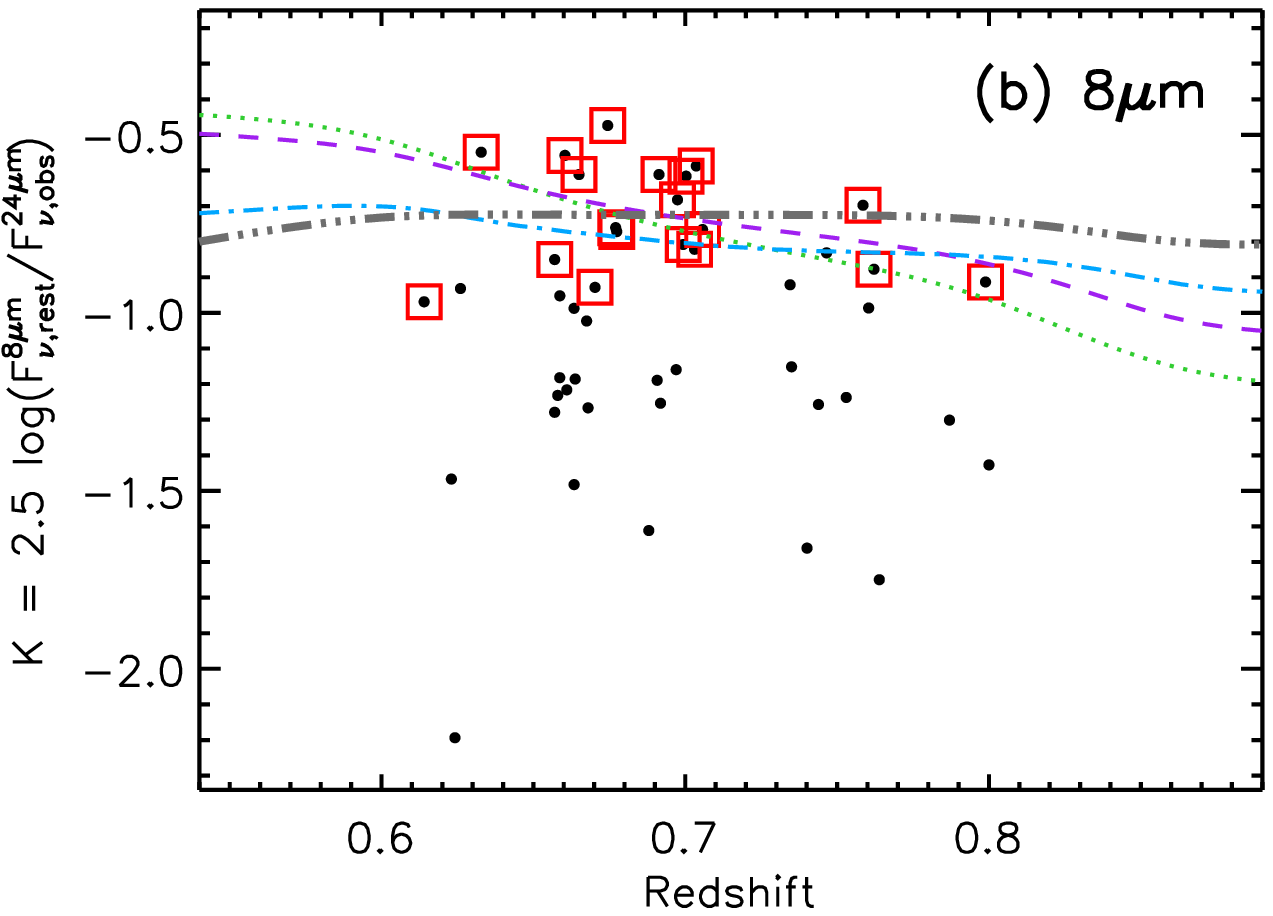}
\caption{($a$) 15\,\um\ and ($b$) IRAC 8\,\um\ \kcors\ from MIPS 24\,\um. The black points show \kcors\ of the IRS sample as measured from their mid-IR spectra. The red boxes highlight the 20 pure star-forming galaxies (\pagn\ = 0). For each of the four IR libraries \citep{Lagache03,Rieke09,Dale02,Chary01}, we show the \kcor\ track of the template that best fit the measured \kcors\ from the 20 star-forming galaxies. The selected templates for \kcors\ are highlighted with thicker lines.
\label{fig:kcorrect}} 
\end{figure*}

Before estimating the LFs, one has to take into account three major sample biases: (1) source extraction incompleteness, (2) optical identification or redshift incompleteness, and (3) photometric redshift uncertainties. Monte Carlo simulations showed that source extraction from the 24\,\um\ mosaic is highly complete; the catalog is about 75\% complete at $F24$ = 60\,\uJy\ and reaches 100\% completeness above 110\,\uJy\ \citep{Le-Floch09}. The completeness of the 70\,\um\ catalog is poor; it is only $\sim$20\% complete at 7 mJy and approaches 100\% completeness above 16 mJy \citep{Frayer09}. The optically unidentified fraction of 24\,\um\ sources at $z\sim0.7$ should be less than 1\%, since only 0.6\% of the MIPS sources between \zbin\ have optical counterparts between $24 <$\,\iab\,$< 26.5$ \citep[Fig.~\ref{fig:completeness}$b$;][]{Le-Floch09}. The photometric redshift catalogs show excellent agreement with spectroscopic redshifts for sources with \iab\,$< 24$ and $z < 1.25$ --- $\sigma_{\Delta z/(1+z)}$ = 0.012(0.015) for normal galaxies and X-ray sources, respectively \citep{Ilbert09,Salvato09}. Since 99.4\% of the 24\,\um\ sources at \zbin\ have optical counterparts brighter than \iab\,$< 24$, photometric redshift scattering should not affect our results. Besides, 29.4\% of the 24\,\um\ sources down to 60\,\uJy\ between \zbin\ have secure spectroscopic redshifts. To conclude, the only sample bias that we need to correct for is from source extraction.

With the IRS spectra, we can compute LFs at any wavelength between 5.5 and 20\,\um. We decided to show LFs at two representative mid-IR wavelengths, 15\,\um\ (monochromatic) and IRAC 8\,\um, for several reasons: (1) 24\,\um\ roughly corresponds to rest-frame 15\,\um\ at $z \sim 0.7$, minimizing bandpass correction, (2) for star-forming galaxies the 15\,\um\ flux is dominated by dust continuum from very small grains stochastically heated by young stars and the IRAC 8\,\um\ filter includes strong PAH features at 7.7 and 8.6\,\um, and (3) local benchmarks at these wavelengths exist in the literature \citep[\eg][]{Xu00,Pozzi04,Huang07}. 

As the luminosity limit of the IRS sample is a few times higher than the predicted ``knee" of the LF, it is necessary to include the entire 24\,\um\ sample to complement these bright-end LFs. Without mid-IR spectra for these sources, we have to {\it K}-correct the observed 24\,\um\ fluxes to rest-frame luminosities using assumed SEDs. Since AGNs are less relevant at lower luminosities (Figs.~\ref{fig:irs} \& \ref{fig:paheqw}), it is reasonable to assume that sources below 0.7 mJy are dominated by star-forming galaxies \citep{Fadda10}. Hence, we used the IRS spectra of SF-dominated galaxies to select the best SED for the \kcor. We compared the measured \kcors\ of the 20 ``pure" star-forming galaxies (\ie\ mid-IR AGN contribution, \pagn, less than 8\%) with predictions of the templates in four commonly used SED libraries for IR galaxies \citep{Chary01,Dale02,Lagache03,Rieke09}. In Figure~\ref{fig:kcorrect}, we show, for each SED library, the track of the template that yielded the smallest residual to the average measured \kcor\ at each redshift. Although we selected the best fit of the four best templates\footnote{A \citet{Rieke09} template (\Lir = $5.3\times10^{10}$ \lsun) at 15 \um\ and a \citet{Lagache03} template (\Lir = $9.1\times10^{12}$ \lsun) at 8 \um.} for subsequently {\it K}-correcting the 24\,\um\ sources, we obtained almost identical results for all four of them. 

We adopted a single template for the \kcor\ because of practical reasons: (1) no trend of \kcors\ relative to luminosity can be detected because of the narrow luminosity range spanned by the pure star-forming galaxies (log(\Lfif/\lsun) = 10.6$-$11.1), and (2) templates that match the observed [8.0]$-$[24] colors at a given redshift produce erroneous \kcors\ at 8 and 15 \um. 

We computed LFs with the 1/\vmax\ method \citep{Schmidt68}. We calculated the lower and upper redshift limits ($z_{\rm min}$ and $z_{\rm max}$) for each galaxy to remain above the MIPS flux limits. Since MIPS sources at $z\sim0.7$ are bright in the optical, the effect of the optical selection (\iab\,$< 26.5$) is negligible. We applied weights (defined as 1/completeness) in the 1/\vmax\ calculation to correct for sample incompleteness below 0.11 and 16 mJy for 24 and 70\,\um, respectively. Errors of the comoving densities were calculated based on Poisson statistics. 

\subsection{Mid-IR Luminosity Functions} \label{MIRLF}

\begin{figure*}[!tb]
\epsscale{0.58}
\plotone{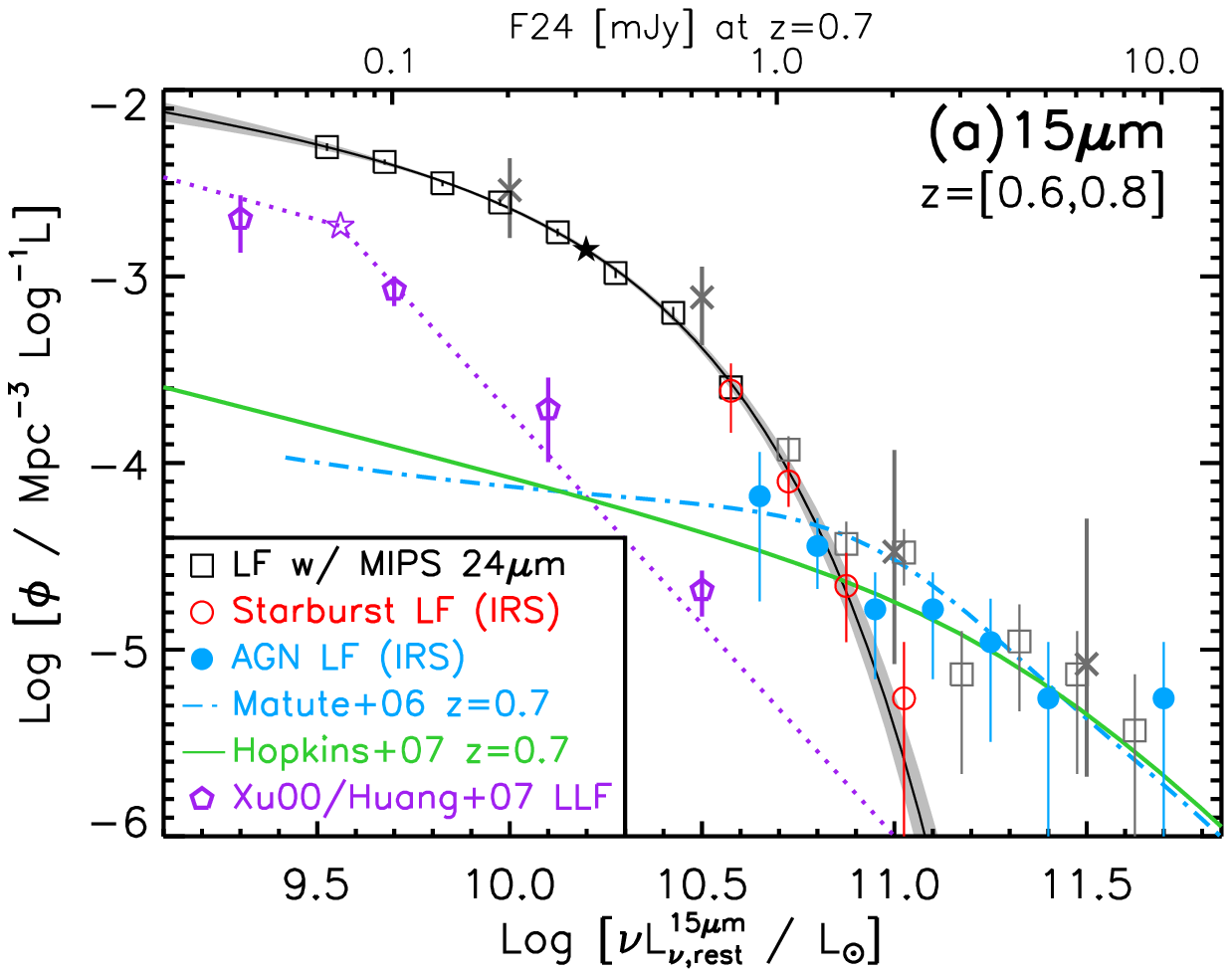}
\plotone{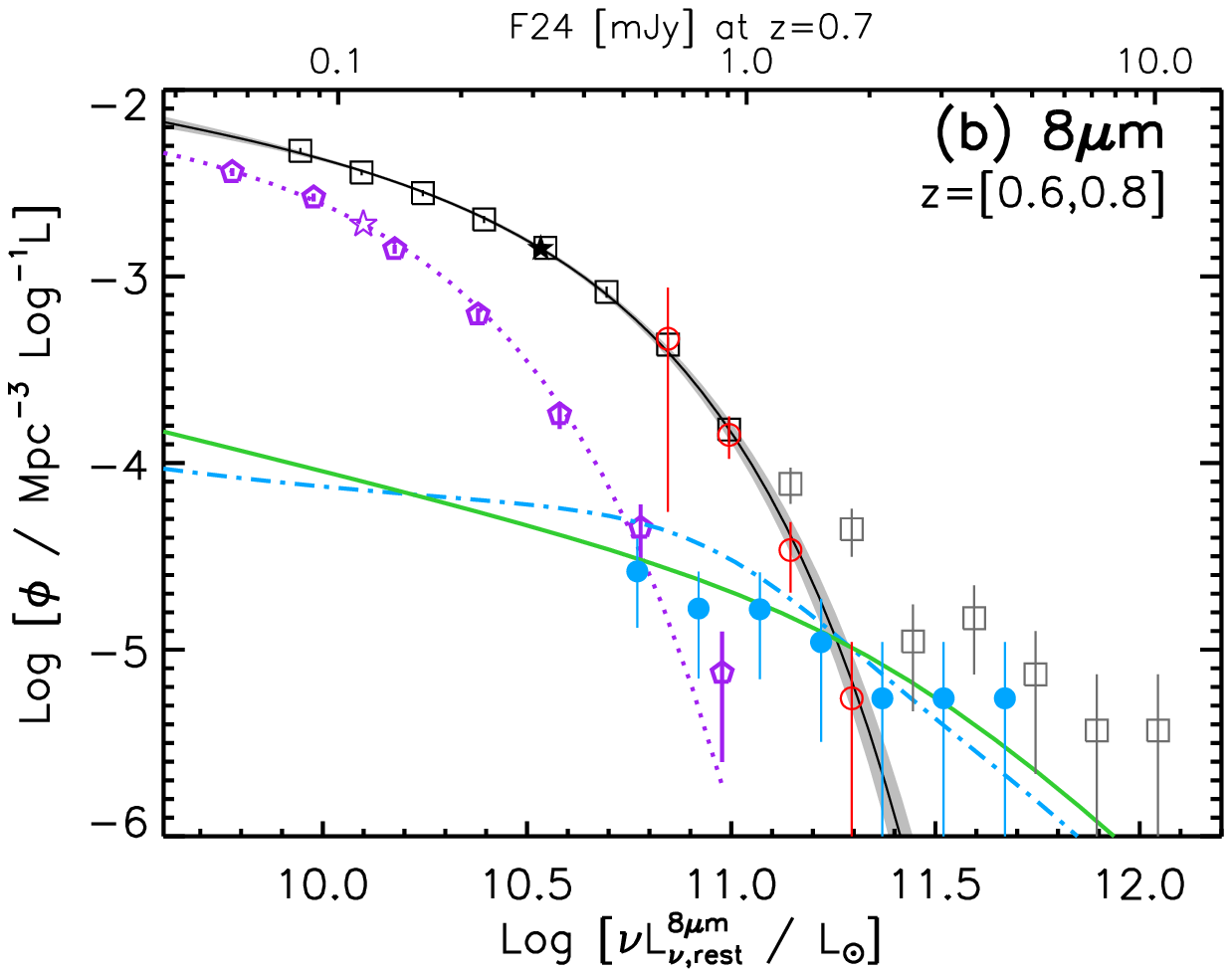}
\caption{($a$) 15\,\um\ and ($b$) IRAC 8\,\um\ LFs at $z \sim 0.7$. The IRS decomposed AGN and SF LFs are shown as blue filled circles and red open circles, respectively. The SF LFs were extended to lower luminosities with the MIPS sample down to $F24$ = 60\,\uJy\ ({\it black/grey squares} show data points below/above the luminosity limit of the IRS sample). The best-fit Schechter functions of the SF LFs are shown as black solid curves, with the shaded areas delimiting the 1$\sigma$ uncertainties from Monte Carlo realizations. As a comparison with previous studies, in the left panel we show the 15\,\um\ LF at \zbin\ of \citet{Le-Floch05} ({\it grey crosses} with error bars). The AGN LFs match the 15\,\um\ LF of optically selected type-1 and type-2 AGNs \citep[{\it blue dashed curve};][]{Matute06} and the {\it K}-corrected obscuration-corrected AGN bolometric LFs \citep[{\it green solid curves};][]{Hopkins07}. To show LF evolution, we included the local 15\,\um\ and 8\,\um\ LFs \citep[{\it purple pentagons} and {\it dotted curves};][]{Xu00,Huang07}. The top abscissa indicates corresponding 24\,\um\ fluxes at $z$ = 0.7. The knees ($L^{\star}$) of the LFs are marked by stars.
\label{fig:LF}} 
\end{figure*}

The total 15\,\um\ LF from the MIPS sample agrees with that of \citet{Le-Floch05}. It is consistent with a modified Schechter profile (Eq. [2]), although our finer luminosity bins revealed a discontinuity at log(\Lfif/\lsun) $\sim$ 10.8. Accounting for the different effective areas, the LFs from the MIPS sample and the IRS sample agree well with each other, confirming that the IRS sample is representative of the complete sample. Using our mid-IR decomposition results (\S~\ref{irs}), we measured AGN and SF luminosities separately for individual sources. The SF and AGN LFs are shown in Fig.~\ref{fig:LF} and are tabulated in Tables~\ref{tab:SFLF} \& \ref{tab:AGNLF}. 

The AGN subtraction removed the relatively large number of high-luminosity sources, which had justified the usage of modified Schechter profiles for the overall mid-IR LFs. As AGNs are unimportant at log(\Lfif/\lsun) $<$ 10.6, we extended the SF LFs to lower luminosities using the LFs from the MIPS sample. Evidently, the SF LFs were well described by Schechter profiles (Eq. [1]), similar to LFs at shorter wavelengths (\S~\ref{intro}). We found the best-fit Schechter function parameters with MPFIT and estimated their uncertainties with 1000 Monte Carlo realizations (Table~\ref{tab:fit}).

We confirmed the strong, almost pure luminosity, evolution in the SF LFs between $0 < z < 0.7$, as one can simply shift the local 15\,\um\ LF by $\sim$0.6 dex in luminosity to fit the total 15\,\um\ LF at $z \sim 0.7$ (Fig.~\ref{fig:LF}). Note that the local 15\,\um\ LF of \citet{Xu00} exhibits a shallow bright-end slope, because the author did not differentiate between AGN and SF. On the other hand, optically selected AGNs were removed from the local ($z < 0.3$) 8\,\um\ LF of \citet{Huang07}, and the authors employed a Schechter function to describe the SF LF. Quantitatively, the integrated 8\,\um\ luminosity density (Table~\ref{tab:fit}) decreased by $\sim$60\% from $z \sim 0.7$ to $z \sim 0.3$. Using the SFR$-$\Leig\ relation of \citet{Huang07}\footnote{Converted to \citet{Chabrier03} IMF.}, SFR = \Leig/($1.78\times10^9$ \lsun) \msun\ yr$^{-1}$, this luminosity density gives a SFR density of $0.043\pm0.021$ \msun\ yr$^{-1}$ Mpc$^{-3}$ at $z \sim 0.7$, where we have assumed a 50\% calibration uncertainty \citep[cf.][]{Wu05,Reddy10}. We defer a further estimate on the SFR density to \S~\ref{IRLF}.

Our AGN LFs agree with the obscuration-corrected AGN bolometric LF \citep{Hopkins07}. The bolometric LFs were determined using observed AGN LFs from X-ray to mid-IR between $0 < z < 6$; obscured AGNs\footnote{Throughout the paper, we use ``type-1" and ``type-2" to describe optical spectral types and we use ``obscured" and ``unobscured" to distinguish whether or not the X-ray absorption column density is greater than $10^{22}$ cm$^{-2}$.} were accounted for using the observed luminosity-dependent AGN absorption distribution \citep{Ueda03} and assuming equal numbers of Compton-thick objects ($N_{\rm H} > 10^{24-25}$ cm$^{-2}$) and AGNs with $N_{\rm H} = 10^{23-24}$ cm$^{-2}$. The model including Compton-thick AGNs provides a better fit to the X-ray background spectrum. We {\it K}-corrected the bolometric LF to rest-frame 8\,\um\ and 15\,\um\ with the luminosity-dependent AGN SED templates of \citet{Hopkins07}. The {\it K}-corrected LFs fit our AGN LFs at both wavelengths remarkably well ({\it green solid curves} in Fig.~\ref{fig:LF}). The AGN contribution is less prominent in the 8\,\um\ LF than in the 15\,\um\ LF because the PAHs increase the SF/AGN contrast.  This result is reassuring, as mid-IR spectroscopy should provide the most complete AGN identification. The obscuration correction factor for the bolometric LF is about 2.7 (\ie\ obscured:unobscured $\simeq$ 1.7:1) in the luminosity range of the IRS sample. The IRS-selected AGNs, therefore, should be predominantly obscured sources. This is confirmed by (1) their low X-ray detection rate (\S~\ref{AGN}), and (2) the rarity of broad emission lines in their optical spectra.

\citet{Matute06} built 15\,\um\ LFs for type-1 AGNs at $z$ = 0.1 and 1.2 and for type-2 AGNs at $z$ = 0.05 and 0.35 using optically selected samples. We evolved their LFs to $z = 0.7$ and combined the number densities of both optical types\footnote{Since all of their evolution models produced similar results in the bright end, we show only the result from the models with evolving faint-end slopes.} ({\it blue dot-dashed curve} in Fig.~\ref{fig:LF}). The result agrees well with our AGN 15\,\um\ LF. Assuming a flat AGN mid-IR SED in $\nu L\nu$, it also fits our AGN 8\,\um\ LF. Even better matches were observed when we kept the emission powered by SF in the AGN/SF composite systems, imitating the situation faced by \citeauthor{Matute06}, who were unable to decompose the mid-IR emission. Nevertheless, since evolving the LFs involved large extrapolation for the type-2 AGN LF, which dominated the AGN number densities, this should be regarded as only tentative evidence that optically selected AGN samples are as complete as mid-IR selected AGN samples at log(\Lfif/\lsun) $\gtrsim 10.5$. 

The comoving space density of AGNs rises above that of star-forming galaxies at log(\Lfif/\lsun) $\gtrsim 10.8$, and it is dominated by type-2 AGNs at log(\Lfif/\lsun) $\lesssim 11.5$. Only six of the 20 IRS-selected AGNs with optical spectra show broad emission lines; the rest only show narrow emission lines. The predominance of type-2 AGNs over star-forming galaxies at high luminosities is supported by the stacked optical spectrum of 24\,\um\ sources at \Lfif\ $\simeq \nu L_{\nu, \rm obs}^{24\mu m} > 10^{11}$ \lsun\ between $0.4 < z < 0.7$, as presented in \citet{Caputi08}. The spectrum shows narrow emission lines and an [O\,{\sc iii}]5007/H$\beta$ ratio of $\sim$3, indicating AGN photoionization. In contrast, the emission-line ratios of 24\,\um\ sources at lower luminosities are more consistent with star-forming regions.   

\subsection{8$-$1000\,\um\ Luminosity Function} \label{IRLF}

\begin{figure}
\epsscale{1.2}
\plotone{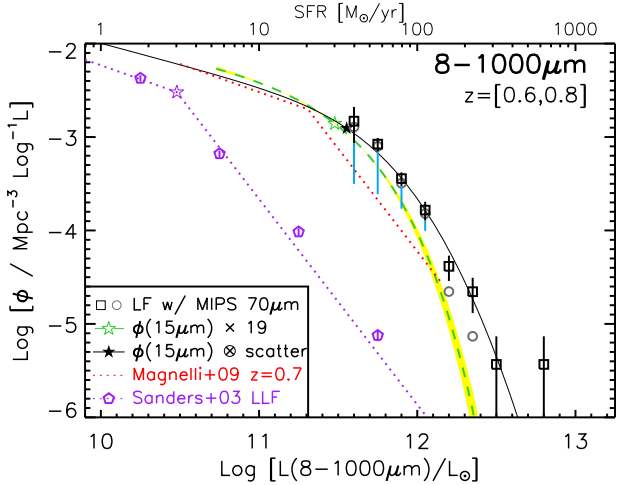}
\caption{8$-$1000\,\um\ LF at $z \sim 0.7$. The 1/\vmax\ data points from the 70\,\um\ selected sample are black squares and grey circles (before and after removing power-law AGNs, respectively). The blue lines extend to the 1/\vmax\ data points before correcting for catalog incompleteness to illustrate the uncertainties below $10^{12}$ \lsun. The green dashed curve is the best-fit SF 15\,\um\ LF, bolometrically corrected adopting the mean \Lir/\Lfif\ ratio of 19 from the stacking results \citep{Lee10}. The solid curve shows the synthetic LF based on the same SF 15\,\um\ LF but incorporated scatter in bolometric correction (see \S~\ref{IRLF} for details). For comparisons, we also included the 8$-$1000\,\um\ LF of \citet{Magnelli09} ({\it red dotted curve}) and the local 8$-$1000\,\um\ LF \citep[purple pentagons;][]{Sanders03}. The knees ($L^{\star}$) of the LFs are marked by stars.
\label{fig:IRLF}} 
\end{figure}

Since the bolometric output of obscured SF emerges mostly as cold dust emission ($T \sim 40$ K), the far-IR wavelengths most directly measure the energetics of active SF, and subsequently, the SFR. Another advantage of the far-IR regime is that the AGN contribution should be minimal, since the average AGN SED drops sharply beyond $\sim$20\,\um. To take a glimpse of the SFR function, we estimated the 8$-$1000\,\um\ LF at $z \sim 0.7$ using a 70\,\um\ selected sample.

Again, we used the 1/\vmax\ estimator. Our sample consisted of 70\,\um-detected sources above 7 mJy between \zbin. The flux threshold is the 3$\sigma$ detection limit for the nominal coverage of 100 for the S-COSMOS field (for a coverage map see \citealt{Frayer09}). The sample area is 1.66 deg$^2$, the same as the 24\,\um\ sample. 

\citet{Symeonidis08} presented a simple equation to convert observed luminosities at 24, 70, and 160\,\um\ to \Lir. This equation\footnote{The $(1+z)$ term must be dropped if one does {\it not} de-redshift the observed flux densities.} produces similar results as more sophisticated SED modeling for a wide redshift range \citep[$\sigma \sim 0.06$ dex;][]{Kartaltepe10}. We thus computed \Lir\ using this equation. For the 70\,\um\ sources undetected at 160\,\um, we assumed an average $F160/F70$ ratio\footnote{Interpolating the stacked $F160/F70$ values according to $F24$ or $F70$ fluxes makes almost no difference in terms of the overall shape of the LF.} of 4, according to the S-COSMOS median stacking analysis \citep{Lee10}. We adopted the completeness curve for coverages greater than the nominal value of 100 to correct for source extraction incompleteness below $F70$ = 16 mJy \citep{Frayer09}. The 1/\vmax\ LF is shown in Figure~\ref{fig:IRLF} and is tabulated in Table~\ref{tab:IRLF}. 

We compared our LF with that of \citet{Magnelli09}\footnote{For comparisons with other determinations of the LF, refer to Fig.~12 of their paper.}, who also used 70\,\um\ data to estimate the bright end of the 8$-$1000\,\um\ LF. We interpolated their LFs to $z = 0.7$. Despite an offset, the bright-end slopes are consistent down to the lowest comoving density accessible to their survey (their largest field is 0.25 deg$^2$). Note that \citeauthor{Magnelli09} fixed the bright-end slope to that of the local 8$-$1000\,\um\ LF \citep[$\phi \propto L^{-2.2}$;][]{Sanders03}. The $\sim$0.2 dex offset is due to the difference in deriving luminosities. The SED templates of \citet{Chary01}, which \citeauthor{Magnelli09} used to compute \Lir, gives \kcors\ (\Lir/$\nu L_{\nu,\rm obs}^{70\mu m}$) 20-40\% lower than those from the equation of \citet{Symeonidis08}. 

\citet{Magnelli09} showed that the 8$-$1000\,\um\ LFs are consistent with the 15\,\um\ LFs once the latter are $K$-corrected with their stacking analysis results. This consistency is expected because both mid-IR and total IR probes obscured SF. This result, however, is moot because their 15\,\um\ LFs are heavily contaminated by AGNs but their far-IR$-$derived 8$-$1000\,\um\ LFs are probably not. Since our SF 15\,\um\ LF has a steeper bright-end slope than the total 15\,\um\ LF, we re-examined the relation between mid-IR and total IR LFs.

\begin{figure}[!tb]
\epsscale{1.2}
\plotone{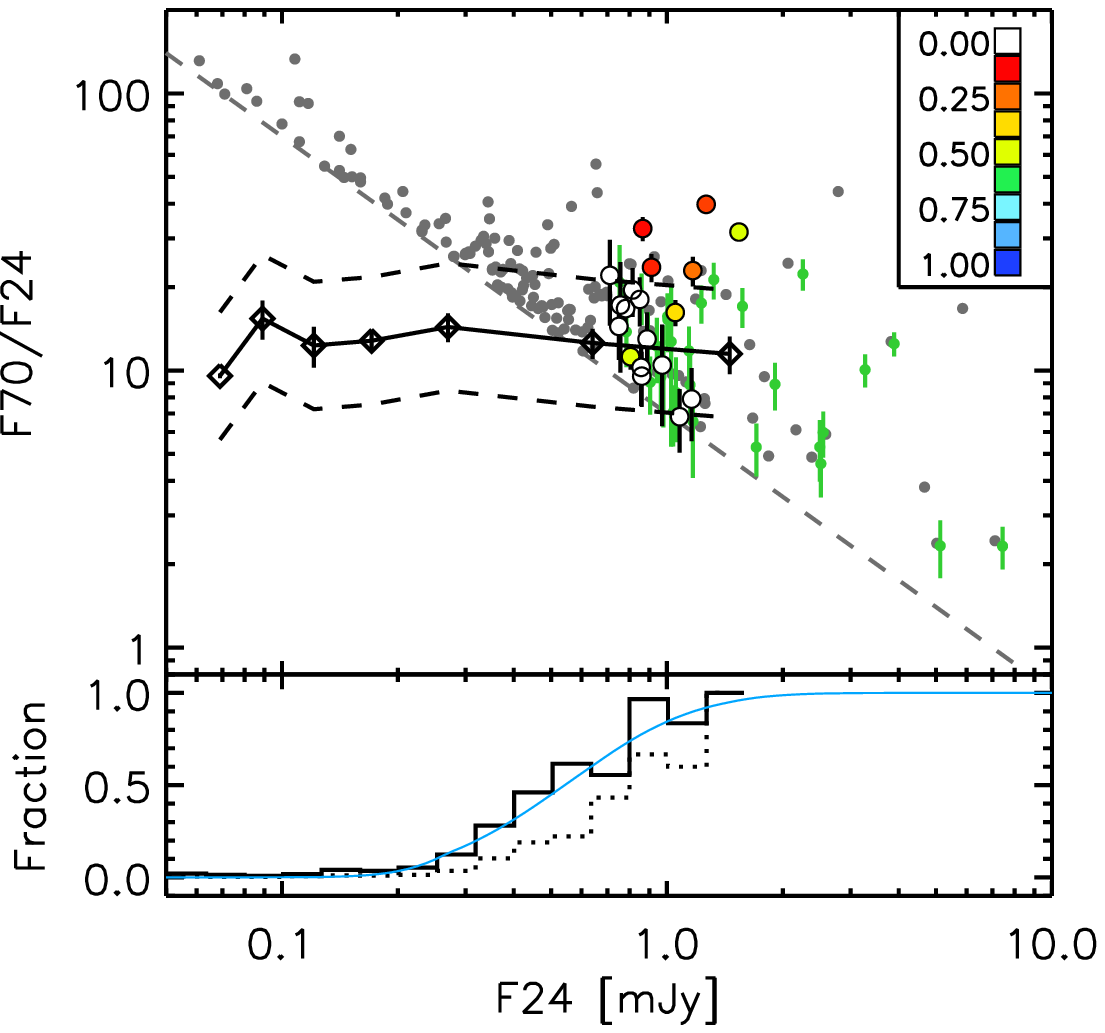}
\caption{70/24\,\um\ flux density ratio vs. 24\,\um\ flux for galaxies at \zbin. Individual detections above 7 mJy are shown as grey filled circles. The green data points highlight the 30 IRS sources detected at 70\,\um; and the big circles show AGN-corrected results for the 19 IRS sources with $F24({\rm SF}) > 0.7$ mJy, color-coded according to the AGN contribution (\pagn; see the legend). The grey dashed line indicates $F70$ = 7 mJy, the 3$\sigma$ detection limit. The stacking analysis results of \citet{Lee10} are shown as black diamonds connected by a solid curve; and the error bars show the expected variation due to \kcor\ within \zbin. In the bottom panel, the solid histogram shows the fraction of 24\,\um\ sources detected at 70\,\um. To show the effect of the catalog incompleteness correction, the dotted histograms show the detection fraction without correcting for this incompleteness. AGN corrections based on IRS spectra were applied for bins above $F24 >$ 0.7 mJy. The blue curve shows the expected fraction of 70\,\um\ detected sources if the $F70/F24$ ratio is a Log-Normal distribution with $\sigma$ = 0.23 dex and means centered on the stacking results. The dashed curves in the upper panel mark this dispersion ($\pm1\sigma$). 
\label{fig:F70F24}} 
\end{figure}

Previous stacking analyses and various luminosity-dependent IR SED libraries have consistently shown that the extrapolation from \Lfif\ to \Lir\ is independent of \Lfif\ or \Lir\ (\eg\ \citealt{Zheng07a,Lee10}; see, \eg\ Fig.~8 of \citealt{Le-Floch05}). Using the average \Lir/\Lfif\ ratio from the S-COSMOS median stacking analysis \citep[$<$\Lir/\Lfif$>$ = 19 at \zbin; Fig.~\ref{fig:F70F24}][]{Lee10}\footnote{We computed \Lir\ for each 24\,\um\ flux bin using the stacking results and the equation of \citet{Symeonidis08}, which gave consistent results as the full SED modeling of \citet{Lee10}; we computed \Lfif\ by {\it K}-correcting the 24\,\um\ fluxes in the same way as in \S~\ref{completeness}.}, we converted the best-fit SF 15\,\um\ LF to an 8$-$1000\,\um\ LF ({\it black dashed curve} in Fig.~\ref{fig:IRLF}). For consistency, we calculated the luminosities of the stacked galaxies with the equation of \citet{Symeonidis08}, although they agreed with those from the SED fitting of \citet{Lee10}. As we suspected from the steep bright-end slope of the 15\,\um\ LF, this {\it K}-corrected LF significantly underestimates the 8$-$1000\,\um\ LF. 

In the {\it median} stacking analysis that we used, the average bolometric conversion factor has included both detections and non-detections at 70/160\,\um. Hence, the observed mismatch is not due to an {\it underestimated} \Lir/\Lfif. In fact, previous {\it mean} stacking analyses gave much smaller conversion factors \citep[$<$\Lir/\Lfif$> \simeq 8$;][]{Zheng07a,Magnelli09}, because 70/160\,\um\ detections were excluded.

An important factor that we have neglected so far is the scatter in the \Lfif$-$\Lir\ correlation. We show the dispersion in \Lir/\Lfif\ by plotting $F70/F24$ against $F24$ (Fig.~\ref{fig:F70F24}). For objects selected within a narrow redshift bin, like ours, the effect of differential \kcor\ is negligible. Since $F70$ strongly correlates with \Lir\ and $F24$ is approximately 15\,\um\ flux at the rest-frame, $F70/F24$ is equivalent to \Lir/\Lfif. The 70\,\um\ detection fraction increases with 24\,\um\ flux, with a profile similar to an error function (bottom panel of Fig.~\ref{fig:F70F24})\footnote{As AGNs dominate above $F24 \gtrsim 1$ mJy, we used the AGN-subtracted IRS sample to calculate the 70\,\um\ detection fraction at $F24 > $ 0.7 mJy.}. Given that \citeauthor{Lee10}'s stacking analysis results represent the average $F70/F24$ for all objects within a given 24\,\um\ flux bin, the increase of 70\,\um\ detection fraction with 24\,\um\ flux can be understood if the $F70/F24$ distribution is Log-Normal with $\sigma$ = 0.23 dex. Admittedly, this measurement is only valid above $F24 \gtrsim 0.3$ mJy, because of the small detection fraction at lower fluxes. But our conclusion is insensitive to the scatter at low 24\,\um\ fluxes.

The dispersion is mostly intrinsic and reflects the variations in IR SEDs of SF. The measurement errors ($\sim$0.1 dex), which include both random errors in the photometry and the systematic calibration uncertainties at both 24 and 70\,\um\ \citep{Engelbracht07,Gordon07}, are low compared to this dispersion. From the stacking analysis results at adjacent redshift bins, we estimated that the spread due to our finite redshift bin is about $\pm$0.05 dex. After subtracting quadratically the measurement uncertainties and the spread due to the redshift bin, we found that the {\it intrinsic} dispersion of the flux density ratio is 0.2 dex. Note that nearby IR galaxies show the same amount of dispersion in the \Lir/\Lfif\ ratios \citep{Chary01}. 

To evaluate the impact of this dispersion to the LF transformation from 15\,\um\ to 8$-$1000\,\um, we generated a mock catalog of galaxies that follows the Schechter profile of the SF 15\,\um\ LF and the redshift distribution of the actual catalog. We then applied inverse \kcors\ to obtain their 24\,\um\ fluxes, and we randomly assigned their 70\,\um\ fluxes using the stacking analysis results and the dispersion of $F70/F24$. The 160\,\um\ fluxes were fixed to the average $F160/F70$ ratio from stacking. Again we calculated \Lir\ using the equation of \citet{Symeonidis08}. As a convolution of a Gaussian and a Schechter function, the synthetic LF is best described by a modified Schechter profile (Eq. [2]), although Schechter function can also provide an acceptable fit. The best-fit model is shown as the black solid curve in Fig.~\ref{fig:IRLF} and its parameters are listed in Table~\ref{tab:fit}. With a shallower bright-end slope, it is now consistent with the 70\,\um$-$derived LF and its extrapolation to $10^{13}$ \lsun\ agrees with that of sub-milimeter galaxies \citep{Chapman05}. 

The synthetic LF also extends the 8$-$1000\,\um\ LF below the ``knee". Integrating the synthetic LF we obtained a total bolometric luminosity density of $7\times10^8$ \lsun\ Mpc$^{-3}$. Using the \citet{Kennicutt98} calibration, SFR = \Lir/($10^{10}$ \lsun) \msun yr$^{-1}$, we obtained a SFR density of $0.07\pm0.02$ \msun\ yr$^{-1}$ Mpc$^{-3}$, where the error is dominated by the $\sim$30\% calibration uncertainties. This result agrees well with previous estimates using non-IR tracers \citep[see the compilation of][]{Hopkins06}. LIRGs/ULIRGs contribute about 54\%/7\% of the total SFR density at $z \sim 0.7$. The ULIRG contribution to the total SFR density would have been underestimated by a factor of four if we had simply converted the SF 15\,\um\ LF using the mean \Lir/\Lfif.

In summary, only after incorporating both the average \Lir/\Lfif\ and its dispersion were we able to correctly predict the 8$-$1000\,\um\ LF from the SF 15\,\um\ LF. Note that our result is \emph{insensitive} to the shape of the \Lir/\Lfif\ distribution: whether or not it is Log-Normal, this mismatch in the bright-end slope can be reconciled as long as $\sim$16\% of the sources with \Lfif\,$\gtrsim 2\times10^{10}$ \lsun\ have \Lir/\Lfif\ ratios 1.6 times higher than the mean value. We suggest two possible physical origins of these galaxies:

\begin{enumerate}

\item These galaxies show the hottest IR SEDs because they are extreme starbursts. It is known that star-forming galaxies form a SFR$-M^{\star}$ sequence with an intrinsic dispersion of $<$0.3 dex \citep{Noeske07a,Zheng07b,Elbaz07,Daddi07a}. Since the SFRs in these studies were derived mostly using 24\,\um\ data, the dispersion in the true SFR$-M^{\star}$ sequence must be larger than the current estimates given the $\sim$0.2 dex dispersion in the ratio of 70\,\um\ and 24\,\um$-$derived SFRs. In analogy, this dispersion can explain why the SFR Function has a much shallower bright-end slope than the stellar Mass Function of star-forming galaxies \citep[][]{Bell07,Ilbert10,Boissier10}. The dispersion could naturally arise if there is a large range of SF efficiency (SFR/$M_{\rm H2}$) \citep{Scoville91,Daddi10} and/or molecular gas fraction ($M_{\rm H2}/M^{\star}$) in star-forming galaxies at a given stellar mass.

\item The rest-frame $\sim$40\,\um\ emission in these galaxies could be powered by AGNs, similar to Mrk\,231 \citep{Veilleux09,Werf10}. In fact, if we remove power-law AGNs ($\alpha^{5}_{2} > -0.2$; \S~\ref{AGN}) from the 70\,\um\ selected sample, then the 8$-$1000 LF becomes more consistent with the SF 15\,\um\ LF as it is significantly reduced at \Lir\,$\gtrsim 10^{12}$ \lsun\ ({\it grey circles} in Fig.~\ref{fig:IRLF}). Although these data points should be considered as lower limits because the IR luminosities in these AGNs could also be powered by obscured SF, it is intriguing that almost all of the IRS sources with $F70/F24$ values 0.23 dex above the stacking results contain AGNs (Fig.~\ref{fig:F70F24}). 

\end{enumerate}

\section{Discussion} \label{discussion}

\subsection{Co-Evolution of BHs and Galaxies in LIRGs} \label{DutyCycle}

Previous multi-wavelength estimates of the AGN and SF LFs have shown that the volume-averaged BH accretion history closely follows the volume-averaged SFR history at $z < 2$ \citep[\eg][]{Shankar09,Zheng09}. The ratio of the two, $\dot{M}_{\rm BH}(z)/{\rm SFR}(z) \simeq 5-8\times10^{-4}$, matches not only the present-day ratio of the BH mass density and stellar mass density ($\rho_{\rm BH}/\rho_{\rm star} \simeq 1.5\times10^{-3}$) but also the normalization of the local $M_{\rm BH}$$-$$M_{\rm bulge}$ relation \citep[$M_{\rm BH}/M_{\rm bulge} \simeq 1.4\times10^{-3}$ at $M_{\rm bulge} = 5\times10^{10}$ \msun;][]{Haring04}, provided that part of the stellar mass is recycled into the interstellar medium. These observations strongly suggest a co-evolution of galaxies and BHs, but are they co-evolving in the same objects and at the same time?

With the AGN/SF spectral decomposition results (\S~\ref{irs}), we can directly address this problem using the 48 galaxies with \Lfif\ $> L_{15}^{\rm lmt} = 3\times10^{10}$ \lsun\ and between \zbin. The 15\,\um\ luminosity limit corresponds to \Lir\ $\simeq 6\times10^{11}$ \lsun\ or SFR $\simeq 60$ \msun\ yr$^{-1}$ for star-forming galaxies and $L_{\rm bol} \simeq 4\times10^{11}$ \lsun\ or $\dot{M}_{\rm BH} = 0.24/f_{0.1}$ \msun\ yr$^{-1}$ for AGNs, where $f_{0.1} = \epsilon/(1-\epsilon)$ for $\epsilon = 0.1$, which is the BH radiation efficiency. Within the sample, 31 sources have SF luminosities ($L_{15}^{\rm SF}$) greater than $L_{15}^{\rm lmt}$, 24 have AGN luminosities ($L_{15}^{\rm AGN}$) greater than $L_{15}^{\rm lmt}$, and 7 have both $L_{15}^{\rm SF}$ and $L_{15}^{\rm AGN}$ above $L_{15}^{\rm lmt}$. 

Intense BH accretion and SF therefore coexist in $\sim$23\% (7/31) of the (U)LIRGs (SFR $> 60$ \msun\ yr$^{-1}$). Assuming every (U)LIRG experiences such an AGN phase in its lifetime, the AGN duty cycle is $0.23\pm0.09$\footnote{Poisson error.} for (U)LIRGs at $z \sim 0.7$. If we assume the observed luminosity distribution evenly samples the light curve of a typical star-forming galaxy or an AGN, then the mass ratio of BH growth and SF can be obtained by integrating luminosities for each component over the 31 galaxies with SFR $>$ 60 \msun\ yr$^{-1}$. The result is $\triangle M_{\rm BH}/\triangle M_{\rm star} \simeq 1.7\times10^{-3}/f_{0.1}$. For the 7 composite galaxies with both components above the luminosity limit, we measured $\sum\dot{M}_{\rm BH}/\sum{\rm SFR} \simeq 4.6\times10^{-3}/f_{0.1}$. Multiplying by the AGN duty cycle (7/31), the time-averaged ratio is $\dot{M}_{\rm BH}/{\rm SFR} \simeq 1.0\times10^{-3}/f_{0.1}$.  

Our estimates are consistent with the local $M_{\rm BH}$$-$$M_{\rm bulge}$ relation \citep{Haring04}, supporting an invariable $M_{\rm BH}$$-$$M_{\rm host}$ relation since $z \sim 1$ \citep{Jahnke09,Bennert10}. For (U)LIRGs, this result also rules out co-evolution scenarios in which the bulges and BHs grow in the same objects and at the same time {\it or} SF and BH accretion occur in different events. Instead, it favors a co-evolution with a time offset between SF and BH accretion, in which BH growth and SF happen in the same event but the former has a much shorter lifetime than the latter.

For the 17 AGNs with $L_{15}^{\rm AGN} > L_{15}^{\rm lmt}$ but $L_{15}^{\rm SF} < L_{15}^{\rm lmt}$, we measured $\sum\dot{M}_{\rm BH}/\sum{\rm SFR} \simeq 1.2\times10^{-2}/f_{0.1}$, almost an order of magnitude higher than the normalization of the local $M_{\rm BH}$$-$$M_{\rm bulge}$ relations. However, since the galaxy comoving density increases dramatically as SFR decreases, the AGN duty cycle in these galaxies is much lower than that of the (U)LIRGs with SFR $>$ 60 \msun\ yr$^{-1}$. With zCOSMOS spectra, \citet{Silverman09} found that on average $\dot{M}_{\rm BH}/{\rm SFR} \simeq 1.9\times10^{-2}$ for X-ray selected AGNs at $z < 1$. These AGN hosts have a mean SFR $\sim 10$ \msun\ yr$^{-1}$. The four times higher $\dot{M}_{\rm BH}/{\rm SFR}$ than what we measured in the composite (U)LIRGs also suggests a smaller AGN duty cycle, probably as a consequence of the longer lifetime for the SF. 

\subsection{Comparison with Other AGN Identification Methods} \label{AGN}

\begin{figure*}[!tb]
\epsscale{0.58}
\plotone{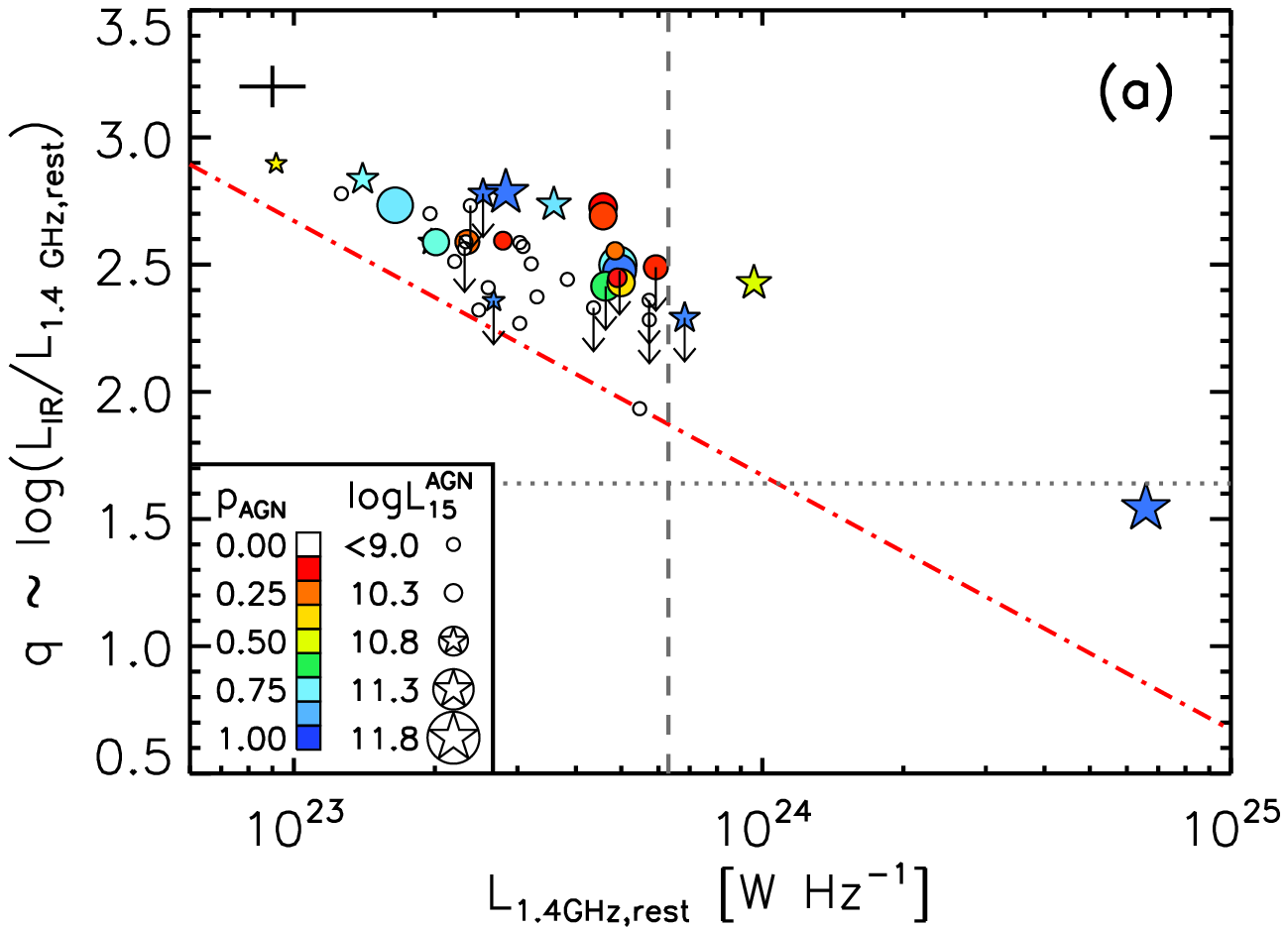}
\plotone{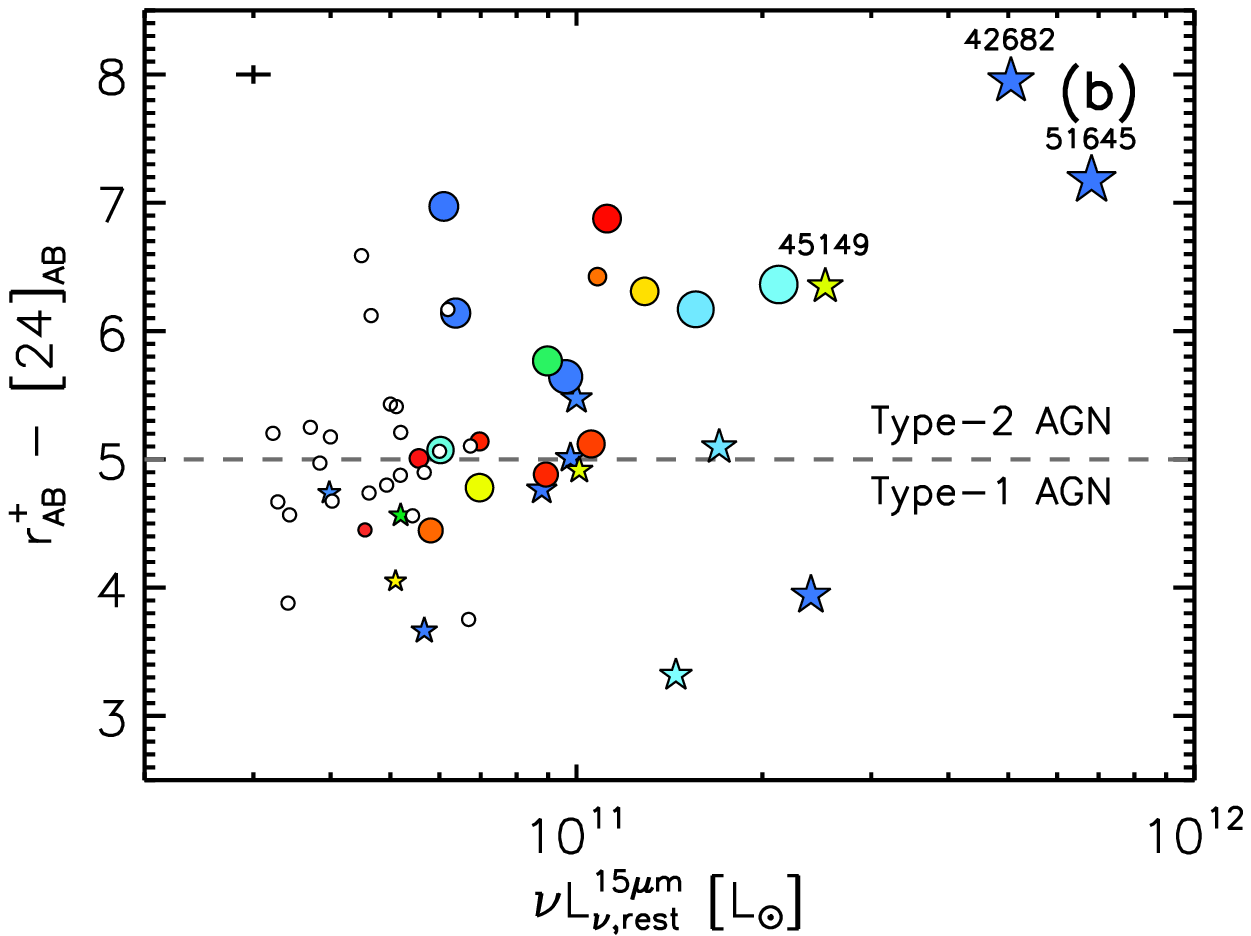}
\plotone{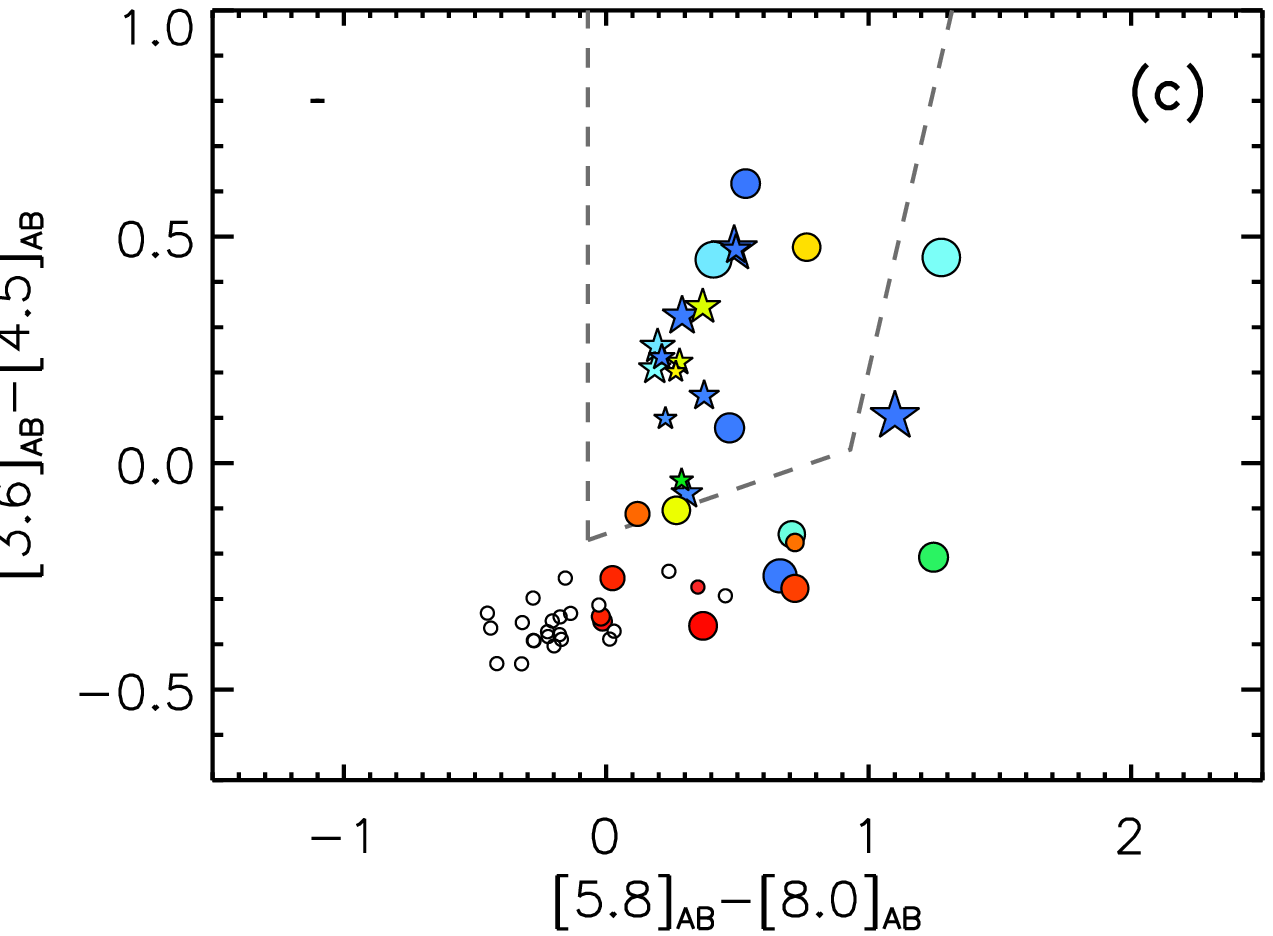}
\plotone{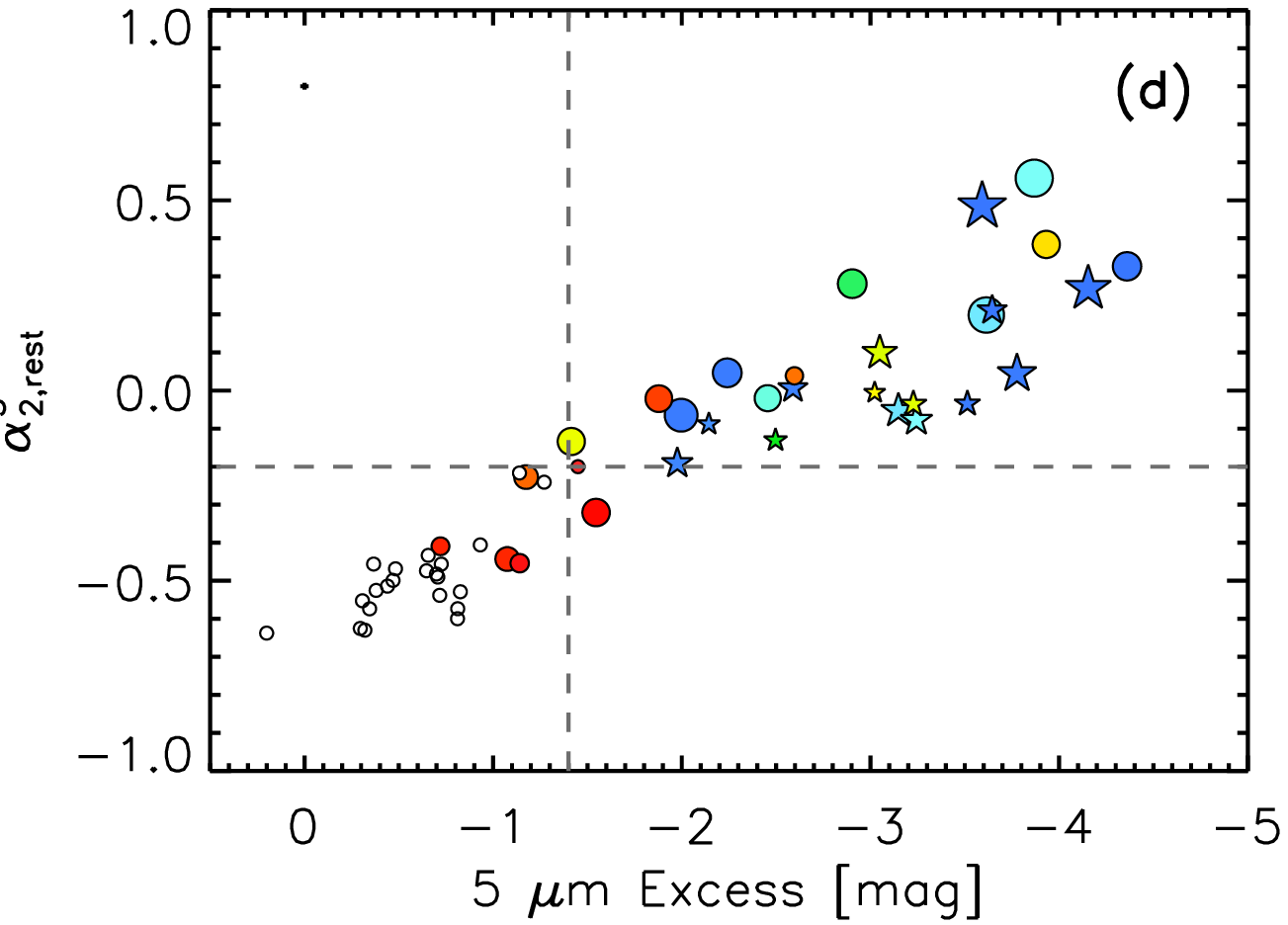}
\caption{Our mid-IR spectroscopical selection of AGNs compared with major AGN identification methods. Symbols are the same as in Fig.~\ref{fig:paheqw}. The median errors are shown at the top left corner in each panel. (a) IR excess ($q$) vs. radio luminosity. Downward arrows mark the radio sources undetected at 70\,\um. The grey dotted and dashed lines indicate the radio selection thresholds ($q < 1.64$ or $L_{\rm 1.4GHz} > 6.3\times10^{23}$ W Hz$^{-1}$; \citealt{Yun01,Hickox09}). The red dot-dashed line shows the IR luminosity limit of the 70\,\um\ sources at $z \sim 0.7$ (\Lir\ $\gtrsim 5\times10^{11}$ \lsun). (b) Optical$-$mid-IR color vs. \Lfif. X-ray AGNs ({\it stars}) show bluer colors (\ie\ less obscured) than non-X-ray AGNs. \citet{Sacchi09}'s Type-1/Type-2 AGN separation line is the dashed line ($F24/F(r) = 100$). The three highly obscured X-ray AGNs are labeled (\S~\ref{CTAGN}). (c) IRAC color-color diagram. The dashed lines delimit the AGN locus of \citet{Stern05}. (d) Rest-frame near-IR spectral index ($\alpha^5_2$ = log($\nu L_{\nu,\rm rest}^{5\mu m}/\nu L_{\nu,\rm rest}^{2\mu m}$)) vs. rest-frame 5\,\um\ excess in magnitude. The 5\,\um\ excess is the magnitude difference between the observed 8\,\um\ flux and the stellar blackbody emission. Most AGNs would be identified if one selects objects with either $\alpha^{5}_{2} > -0.2$ or 5\,\um\ excess $< -1.4$ mag. 
\label{fig:AGN}} 
\end{figure*}

Mid-IR spectral analysis offers arguably the best method to identify AGNs. In most cases, however, one has to rely on photometric data to select AGNs, since it is difficult to carry out mid-IR spectroscopy for a large number of sources. We compared our mid-IR decomposition results with commonly used photometry-based AGN identification methods to select the most efficient methods for future studies. We emphasize that since our comparisons are limited to the IRS sample, \ie\, 24\,\um\ sources above 0.7\,mJy and between \zbin, we cannot comment on the contaminations by sources below the 24\,\um\ flux limit and/or at other redshifts.

From our spectral decomposition of the 53 IRS sources\footnote{These include the 5 objects outside of the redshift bin, \zbin.}, we estimated both the AGN luminosity and its contribution to the total mid-IR SED. For 31 sources, a power-law AGN was required to fit the mid-IR spectrum. In four such sources, the AGN luminosity ($L_{15}^{\rm AGN}$) fell below $3\times10^{10}$ \lsun\ ($L_{15}^{\rm lmt}$). Although these four sources are possibly real AGNs, we ignored them in this discussion because the sample is incomplete below the luminosity limit. We thus have 27 ``AGN hosts" with $L_{15}^{\rm AGN} \gtrsim L_{15}^{\rm lmt}$ and 21 ``AGN-dominated sources" with \pagn\ $>$ 50\%. All of the AGN-dominated sources are also AGN hosts. 

41 IRS sources have radio counterparts from the VLA-COSMOS survey \citep{Schinnerer07}. Three can be classified as radio AGNs (Fig.~\ref{fig:AGN}$a$), as their rest-frame radio luminosities are greater than $6.3\times10^{23}$ W Hz$^{-1}$ \citep[\eg][]{Hickox09}. An alternative selection method is based on the IR-radio correlation \citep{Helou85}. \citet{Yun01} suggested that sources with $q$ parameters five times less than the mean value (\ie\ $q < 1.64$) are AGNs, where $q$ = log(\Lir/\lsun)$-$log(L$_{\rm 1.4GHz}$/$10^{14}$ W Hz$^{-1}$). Our most radio luminous source is also the only radio-excess object with $q = 1.54$ \citep[see][]{Sargent10}. We calculated \Lir\ in the same way as in \S~\ref{IRLF}. The 11 radio sources without 70\,\um\ detection are marked by downward arrows. Radio power/radio excess identified only 11/4\% (3/27,1/27) of the AGN hosts.

14 sources have X-ray counterparts from the XMM-COSMOS survey \citep[{\it stars} in Fig.~\ref{fig:AGN};][]{Hasinger07,Brusa10}, including the 3 radio-selected AGNs. These are bona fide AGNs, since their rest-frame X-ray luminosities ($L_{\rm X} = L_{\rm 0.5-10keV}$) are all greater than $10^{43}$ \ergs\ given a photon index of $\Gamma = 1.7$. The X-ray selection, however, identified only 52\% (14/27) of the AGN hosts and 62\% (13/21) of the AGN-dominated sources. The central 0.9 deg$^2$ of the COSMOS field was covered by the much deeper \chandra\ COSMOS survey (\citealt{Elvis09}, Civano et al. 2010, in prep.). The \chandra\ detection limit was $f_{\rm 0.5-10 keV} = 5.7\times10^{-16}$ \ergs\ cm$^{-2}$, corresponding to $L_{\rm X} = 10^{42}$ \ergs\ at $z \sim 0.7$. Eight XMM-selected AGNs were inside the \chandra\ coverage, and they were all in the \chandra\ catalog with consistent fluxes. Additionally, \chandra\ detected two pure star-forming galaxies (\pagn\ = 0) with $L_{\rm X} \lesssim 1.5\times10^{42}$ \ergs\ that were not in the XMM catalog. We attributed the X-ray emission of these two objects to SF activity. Although 11 of the 13 non-X-ray AGNs were observed by \chandra, none was detected.

The \spitzer/IRAC-based AGN diagnostics select sources with power-law SEDs, presumably dominated by AGN heated hot dust emission. The IRAC color-color selection of \citet{Stern05} identified 67\% (18/27) of the AGN hosts and 71\% (15/21) of the AGN-dominated sources (Fig.~\ref{fig:AGN}$c$). The near-IR spectral index between rest-frame 2 and 5\,\um, $\alpha^{5}_{2}$ = log($L_5/L_2$), is even more efficient. Since the effective wavelengths of the IRAC 3.6 and 8.0\,\um\ filters correspond to rest-frame 2.1 and 4.5\,\um\ at $z \sim 0.7$, we computed $\alpha^{5}_{2}$ using [3.6] and [8.0] magnitudes directly. Applying $\alpha^{5}_{2} > -0.2$, we identified 89\% (24/27) of the AGN hosts and 100\% (21/21) of the AGN-dominated sources. The rest-frame 5\,\um\ excess selection works equally well, as indicated by its tight correlation with $\alpha^{5}_{2}$ (Fig.~\ref{fig:AGN}$d$). We computed the excess by subtracting the extrapolated stellar blackbody emission from the observed IRAC 8\,\um\ magnitude. The extrapolation used the best-fit \citet{Bruzual03} model to the $FUV$-to-$K_S$ SED (\S~\ref{irs}). The 2-to-5\,\um\ spectral index and the rest-frame 5\,\um\ excess are, therefore, the best photometric AGN diagnostics, as their results are most consistent with the mid-IR spectral decomposition results.

\subsection{Compton Thick AGNs?} \label{CTAGN}

The remarkable agreement between mid-IR LFs of IRS-selected AGNs and obscuration-corrected AGN bolometric LFs implies that we may have achieved a complete AGN identification (\S~\ref{MIRLF}; Fig.~\ref{fig:LF}). We have confidence in this result because of our high spectroscopic redshift completeness (79\%; \S~\ref{irsobs}) for X-ray$-$detected bright 24\,\um\ sources ($F24 > 0.7$\,mJy). Since most unobscured AGNs with $L_{15}^{\rm AGN} > L_{15}^{\rm lmt}$ should be detected in X-ray at $z \sim 0.7$ and most obscured AGNs should not have strange optical SEDs, the redshifts of the AGNs in the parent sample should be accurate. Recall that Compton-thick AGNs were counted in the bolometric LFs, using the $N_{\rm H}$ distribution of \citet{Ueda03} and assuming as many Compton-thick objects as AGNs with $N_{\rm H} = 10^{23-24}$ cm$^{-2}$ \citep{Hopkins07}. Following the same $N_{\rm H}$ distribution, we estimated that roughly 35\%, 50\%, and 15\% of the IRS-selected AGNs were unobscured, Compton-thin ($N_{\rm H} = 10^{22-24}$ cm$^{-2}$), and Compton-thick, respectively. 

The optical-to-24\,\um\ color has been widely used to select obscured AGNs \citep{Fiore08,Fiore09,Dey08}. Our non-X-ray AGNs appear more obscured than X-ray AGNs as they show redder $r^+_{\rm AB}-[24]_{\rm AB}$ colors; and the two groups are roughly separated at $F24/F(r) = 100$, the type-1/type-2 dividing line suggested by \citet{Sacchi09} (Fig~\ref{fig:AGN}$b$). Since six of the 11 X-ray AGNs and zero of the nine non-X-ray AGNs with optical spectra show broad emission lines, we could further assume that most of the unobscured AGNs were detected in X-ray. Hence, only $\sim$23\% of the obscured AGNs are detected in X-ray. Are there any Compton-thick ones in these X-ray AGNs? The three X-ray AGNs with the highest 15\,\um\ luminosities (MIPS\,42682, 45149, \& 51645; see the last three panels in Fig.~\ref{fig:irs}) show the reddest $F24/F(r)$ colors as well as the highest X-ray hardness ratios, HR = $(H-S)/(H+S) > 0.5$ (others have HR $<$ 0), where $H$ and $S$ are the counts in the $2-10$ and $0.5-2$ keV energy bands, respectively. Two of the three have optical spectra, and both show only narrow emission lines. Although unlikely Compton-thick, these sources are heavily obscured quasars, as HRs above 0.5 indicate $N_{\rm H} > 10^{23}$ cm$^{-2}$ and their absorption-corrected X-ray luminosities are much greater than $10^{44}$ \ergs\ \citep{Mainieri07}. In fact, MIPS\,45149 (XID = 70, HR = 0.7) was included in \citeauthor{Mainieri07}'s X-ray spectral analysis; they modeled its X-ray spectrum using an absorbed power-law with $N_{\rm H} = 1.7\times10^{23}$ cm$^{-2}$ and an intrinsic rest-frame luminosity of $L_{\rm 0.5-10 keV}^{\rm intrinsic} = 6.8\times10^{44}$ \ergs. In the IRS sample, MIPS\,45149 is also the most heavily extinct X-ray AGN in the mid-IR ($\tau_{9.7}^{\rm AGN} = 2.1$).

It is natural to suspect that the four objects with the highest mid-IR extinctions for the AGN component (MIPS\,30035, 48730, 49150, 51523), $\tau_{9.7}^{\rm AGN}$ = 3.4$-$5, are the Compton-thick sources (see Fig.~\ref{fig:irs}), since (1) all of the four were {\it undetected} in X-ray, (2) 80\% (21/27) of the IRS-selected AGNs have $\tau_{9.7}^{\rm AGN} < 1$, and (3) the highest $\tau_{9.7}^{\rm AGN}$ measured in the X-ray AGNs is only 2.1. Note, however, that these optical depths correspond to only $N_{\rm H}$ = 1$-$1.6$\times10^{23}$ cm$^{-2}$ if one assumes the Galactic dust-to-gas ratio and the extinction law of \citet{Rieke85}. By definition, without X-ray data on these objects, we are unable to classify these sources as Compton-thick AGNs.

\section{Conclusions} \label{conclusion}

The mass assembly history of galaxies and their BHs are imprinted in time-resolved IR LFs. Our understanding of both processes have been hindered by the degeneracy of BH accretion and obscured SF in the LFs, especially in the mid-IR regime. In this paper we used \spitzer\ mid-IR spectra to break this degeneracy and decompose mid-IR LFs between SF and AGNs at $z \sim 0.7$. Further, we built a 8$-$1000\,\um\ LF with 70\,\um\ data to evaluate the high-end of the SFR function and to examine the relation between mid-IR LFs and the SFR Function. Finally, we discussed the implications of our analysis to BH-galaxy co-evolution scenarios, AGN selection methods, and the elusive Compton-thick AGN population. Our conclusions can be summarized as follows: 

\begin{enumerate}
\item The SF mid-IR LFs have steep bright-end slopes. Their profiles are well described by the classic Schechter function over three decades in comoving density, similar to LFs at shorter wavelengths. 

\item The AGN mid-IR LFs have shallower bright-end slopes. AGNs are thus responsible for the bright-end excess of mid-IR LFs. The AGN mid-IR LFs, once {\it K}-corrected, match the obscuration-corrected AGN bolometric LF, which was determined with the observed AGN multi-wavelength LFs after accounting for obscured AGNs (both Compton-thin and Compton-thick) with the AGN absorption distribution. The $N_{\rm H}$ distribution predicts that 15\% ($\sim$4) of the AGNs in our IRS sample are Compton-thick. 

\item Although our AGNs are mid-IR selected, the LFs are also consistent with extrapolations from previous determinations with optically-selected AGNs. This tentatively implies that optical spectroscopy can identify most of the obscured AGNs, at least for those with high IR luminosities.

\item The 8$-$1000\,\um\ LF derived from a 70\,\um\ selected sample shows a shallower bright-end slope than the bolometrically-corrected SF 15\,\um\ LF. The mismatch can be reconciled only after incorporating the intrinsic dispersion in the $L_{15}$$-$$L_{\rm IR}$ correlation, especially the $\sim$16\% of 24\,\um\ sources with high $L_{\rm IR}/L_{15}$ ratios ($>1.6\times$ higher than the mean). These sources could be either extreme starbursts or SF/AGN composite systems in which bulk of the far-IR emission is powered by BH accretion. We further suggest that the dispersion in the SFR$-M^{\star}$ sequence explains why the SFR Function has a much shallower bright-end-slope than the stellar Mass Function of star-forming galaxies and the dispersion naturally arises if there is a large range of SF efficiency (SFR/$M_{\rm H2}$) and/or molecular gas fraction in star-forming galaxies at a given stellar mass. 

\item Intense BH accretion accompanies about a quarter of a LIRG's lifetime when the SFR soars above 60 \msun\ yr$^{-1}$. Throughout the LIRG's lifetime, the growth in BH mass is $\sim$0.1\% of the mass of newly formed stars, consistent with the local $M_{\rm BH}$$-$$M_{\rm bulge}$ relation, provided that part of the stellar mass is recycled into the interstellar medium. These results support a constant $M_{\rm BH}$$-$$M_{\rm host}$ relation since $z \sim 1$ and favor co-evolution scenarios in which BH growth and SF are triggered in the same event but the former spans a much shorter lifetime than the latter.

\item X-ray selection ($L_{\rm X} > 10^{42}$ \ergs) misses half of the mid-IR identified AGNs. The majority of the X-ray AGNs are unobscured, despite that the three most mid-IR-luminous X-ray AGNs appear highly obscured ($N_{\rm H} > 10^{23}$ cm$^{-2}$). The majority of the non-X-ray AGNs are obscured. We suspect that the four sources with the highest silicate extinctions, as measured in their power-law components ($\tau_{9.7}^{\rm AGN}$ = 3.4$-$5), are Compton-thick AGNs. All of the four are {\it undetected} in X-ray.

\item Among the photometric AGN identification methods, near-IR spectral index and IRAC excess are most consistent with our mid-IR spectral decomposition analysis.
\end{enumerate}

Although our results are limited to a narrow redshift bin (\zbin), the above conclusions could be extended to other redshifts. The evolution of IR LFs at $0 < z < 3$ based on \spitzer\ and {\it Herschel} data will be the subject of a future paper (Le~Floc'h et al. 2010, in prep.). Extending the AGN/SF decomposition of IR LFs to lower luminosities and for a wider redshift range will greatly improve the measurements of the growth rates of galaxies and their BHs. Together with better constraints on the redshift and mass-dependent normalization in the mass$-$growth-rate relations (\ie\ the specific SFRs and the BH Eddington ratios), we will finally be able to retrace the evolutionary paths of galaxies and their BHs and to witness how the Magorrian relation emerged. 

\acknowledgments 
We thank Knud Jahnke, Mark Sargent, Giovanni Zamorani, and the anonymous referee for helpful comments. This work is based on observations made with the {\it Spitzer} Space Telescope, which is operated by the Jet Propulsion Laboratory, California Institute of Technology, under a contract with NASA. Support for this work was provided by NASA through an award issued by JPL/Caltech under grant JPL-1344606.

\clearpage


\clearpage
\begin{deluxetable}{lccccrrrcccc}
\setlength{\tabcolsep}{.11cm}
\tablewidth{0pt}
\tabletypesize{\footnotesize}
\tablecaption{\spitzer\ IRS Sample \label{tab:sample}}
\tablehead{ 
\colhead{MIPS ID} & \colhead{Opt. RA} & \colhead{Opt. Dec} & \colhead{$z$} & \colhead{$z$flag} & 
\colhead{SL1} & \colhead{LL2} & \colhead{LL1}  &
\colhead{$F24$} & \colhead{\Lfif} &\colhead{\Leig} & \colhead{\pagn} 
\\
\colhead{(1)} & \colhead{(2)} & \colhead{(3)} & \colhead{(4)} & \colhead{(5)} & 
\colhead{(6)} & \colhead{(7)} & \colhead{(8)} & 
\colhead{(9)} & \colhead{(10)} & \colhead{(11)} & \colhead{(12)}
}
\startdata
\hline
 2176 &149.91600&1.879972&0.834&spec-$z$&$20\times2\times 60$&$15\times2\times120$&$10\times2\times120$&0.74&10.65&11.06&0.00\\
 5264 &149.80170&2.384056&0.704&spec-$z$&$13\times2\times 60$&$11\times2\times120$&$ 7\times2\times120$&0.86&10.54&11.06&0.00\\
15123 &150.05206&2.126712&0.665&spec-$z$&$10\times2\times 60$&$10\times2\times120$&$ 6\times2\times120$&0.89&10.57&11.00&0.00\\
16913 &149.87215&2.289672&0.700&spec-$z$&$12\times2\times 60$&$13\times2\times120$&$ 8\times2\times120$&0.78&10.58&10.97&0.00\\
18902*&149.75092&2.469942&0.657&spec-$z$&$ 5\times2\times 60$&$ 9\times2\times120$&$ 7\times2\times120$&0.79&10.72&10.74&0.64\\
18984 &150.15953&2.474338&0.691&spec-$z$&$15\times2\times 60$&$14\times2\times120$&$ 7\times2\times120$&0.79&10.69&11.01&0.00\\
20765 &150.04402&2.643906&0.697&spec-$z$&$ 8\times2\times 60$&$ 7\times2\times120$&$ 5\times2\times120$&1.17&10.75&11.15&0.00\\
23930 &149.91074&1.662555&0.633&spec-$z$&$ 8\times2\times 60$&$12\times2\times120$&$ 8\times2\times120$&0.75&10.52&10.90&0.00\\
26741 &149.71869&1.849873&0.677&spec-$z$&$12\times2\times 60$&$12\times2\times120$&$ 6\times2\times120$&0.85&10.73&10.99&0.00\\
30035 &150.21017&2.311710&0.747&spec-$z$&$ 8\times2\times 60$&$10\times2\times120$&$ 6\times2\times120$&0.87&10.75&11.06&0.11\\
31181*&150.63902&2.464456&0.800&spec-$z$&$ 4\times2\times 60$&$ 7\times2\times120$&$ 6\times2\times120$&0.87&11.00&10.91&0.98\\
31635 &150.26385&2.541551&0.712&spec-$z$&$11\times2\times 60$&$11\times2\times120$&$ 6\times2\times120$&0.81&10.83&10.98&0.00\\
36942 &150.42406&2.123016&0.699&spec-$z$&$15\times2\times 60$&$14\times2\times120$&$ 7\times2\times120$&0.75&10.60&10.97&0.00\\
38331 &149.60471&2.452618&0.674&spec-$z$&$ 8\times2\times 60$&$10\times2\times120$&$ 6\times2\times120$&0.82&10.53&11.01&0.00\\
38754 &149.73370&2.575010&0.706&spec-$z$&$ 6\times2\times 60$&$ 9\times2\times120$&$ 5\times2\times120$&0.81&10.70&10.96&0.00\\
40291 &150.10332&1.661929&0.834&spec-$z$&$17\times2\times 60$&$14\times2\times120$&$ 7\times2\times120$&0.77&10.79&11.12&0.00\\
40410*&150.24252&1.766390&0.623&spec-$z$&$ 4\times2\times 60$&$ 8\times2\times120$&$ 6\times2\times120$&0.73&10.60&10.54&0.97\\
42682*&149.58641&1.769320&0.787&spec-$z$&$ 2\times2\times 60$&$ 2\times2\times120$&$ 2\times2\times120$&5.02&11.70&11.64&1.00\\
42997 &150.66016&1.863509&0.624&spec-$z$&$25\times2\times 60$&$14\times2\times120$&$ 9\times2\times120$&1.05&10.78&10.39&0.74\\
43504 &149.99924&2.005987&0.760&spec-$z$&$ 3\times2\times 60$&$ 4\times2\times120$&$ 4\times2\times120$&1.42&11.03&11.22&0.27\\
44180 &150.21385&2.188338&0.799&spec-$z$&$15\times2\times 60$&$15\times2\times120$&$ 8\times2\times120$&0.86&10.83&11.13&0.00\\
44883 &149.94170&2.395768&0.759&spec-$z$&$10\times2\times 60$&$12\times2\times120$&$ 7\times2\times120$&0.76&10.72&11.04&0.00\\
45149*&150.15022&2.475196&0.692&spec-$z$&$ 2\times2\times 60$&$ 2\times2\times120$&$ 2\times2\times120$&3.82&11.40&11.40&0.54\\
46960 &149.72560&1.810861&0.747&spec-$z$&$10\times2\times 60$&$ 7\times2\times120$&$ 5\times2\times120$&1.23&11.05&11.27&0.14\\
48730 &150.18233&1.700827&0.735&spec-$z$&$ 2\times2\times 60$&$ 2\times2\times120$&$ 2\times2\times120$&1.79&11.11&11.29&0.41\\
48760*&150.68904&1.757104&0.586&phot-$z$&$ 2\times2\times 60$&$ 2\times2\times120$&$ 2\times2\times120$&2.13&10.94&10.90&0.99\\
49150 &150.09402&2.299131&0.691&spec-$z$&$11\times2\times 60$&$11\times2\times120$&$ 5\times2\times120$&0.87&10.76&10.79&0.25\\
49207 &149.96306&2.431469&0.663&phot-$z$&$ 5\times2\times 60$&$ 8\times2\times120$&$ 5\times2\times120$&1.09&10.80&10.73&0.99\\
49223 &150.32326&2.449379&0.744&spec-$z$&$ 6\times2\times 60$&$ 8\times2\times120$&$ 5\times2\times120$&0.94&10.95&10.95&0.67\\
49946*&150.63387&2.593702&0.658&spec-$z$&$ 2\times2\times 60$&$ 2\times2\times120$&$ 2\times2\times120$&2.16&11.16&11.17&0.76\\
50513*&149.65569&2.600808&0.735&spec-$z$&$ 2\times2\times 60$&$ 4\times2\times120$&$ 3\times2\times120$&1.18&10.99&11.00&0.99\\
50519 &150.35817&2.638588&0.703&spec-$z$&$10\times2\times 60$&$11\times2\times120$&$ 7\times2\times120$&0.83&10.72&11.01&0.00\\
50717 &149.68130&2.681122&0.659&spec-$z$&$ 9\times2\times 60$&$ 8\times2\times120$&$ 5\times2\times120$&1.19&10.84&10.88&0.17\\
51182*&150.62514&1.802925&0.626&spec-$z$&$ 2\times2\times 60$&$ 2\times2\times120$&$ 2\times2\times120$&3.04&11.23&11.35&0.85\\
51359 &149.79614&1.774491&0.740&phot-$z$&$ 5\times2\times 60$&$ 5\times2\times120$&$ 4\times2\times120$&1.17&10.98&10.81&0.99\\
51415 &149.69719&1.905216&0.663&spec-$z$&$ 6\times2\times 60$&$10\times2\times120$&$ 6\times2\times120$&0.84&10.66&10.82&0.09\\
51416 &150.20341&1.902644&0.753&spec-$z$&$ 4\times2\times 60$&$ 2\times2\times120$&$ 2\times2\times120$&2.59&11.33&11.32&0.76\\
51427 &150.02583&1.926421&0.661&spec-$z$&$ 4\times2\times 60$&$ 2\times2\times120$&$ 2\times2\times120$&2.38&11.19&11.18&0.83\\
51429 &149.62653&1.931427&0.664&spec-$z$&$ 4\times2\times 60$&$ 5\times2\times120$&$ 4\times2\times120$&1.22&10.84&10.86&0.52\\
51477 &150.48152&2.011073&0.660&spec-$z$&$ 6\times2\times 60$&$ 7\times2\times120$&$ 5\times2\times120$&1.15&10.71&11.18&0.00\\
51489 &149.87846&2.031983&0.677&spec-$z$&$ 5\times2\times 60$&$ 7\times2\times120$&$ 5\times2\times120$&1.08&10.67&11.00&0.00\\
51497*&150.44370&2.049112&0.668&spec-$z$&$ 4\times2\times 60$&$ 7\times2\times120$&$ 6\times2\times120$&0.88&10.75&10.74&0.99\\
51523 &150.43016&2.086895&0.659&spec-$z$&$ 4\times2\times 60$&$ 2\times2\times120$&$ 2\times2\times120$&2.06&11.02&11.24&0.21\\
51538 &149.76747&2.117411&0.670&spec-$z$&$11\times2\times 60$&$10\times2\times120$&$ 5\times2\times120$&0.97&10.60&10.90&0.00\\
51604*&149.91245&2.200366&0.688&spec-$z$&$ 4\times2\times 60$&$ 2\times2\times120$&$ 2\times2\times120$&1.84&11.00&10.90&0.53\\
51645*&149.57237&2.262714&0.764&phot-$z$&$ 2\times2\times 60$&$ 2\times2\times120$&$ 2\times2\times120$&7.12&11.83&11.59&1.00\\
51653 &149.80357&2.288883&0.762&spec-$z$&$ 8\times2\times 60$&$ 8\times2\times120$&$ 4\times2\times120$&1.16&10.78&11.05&0.00\\
51728 &150.28973&2.400017&0.614&spec-$z$&$14\times2\times 60$&$15\times2\times120$&$ 9\times2\times120$&0.71&10.51&10.68&0.00\\
51739*&150.47203&2.410231&0.668&spec-$z$&$ 4\times2\times 60$&$ 5\times2\times120$&$ 4\times2\times120$&1.12&10.71&10.93&0.48\\
51852*&150.05379&2.589671&0.697&spec-$z$&$ 2\times2\times 60$&$ 2\times2\times120$&$ 2\times2\times120$&3.39&11.38&11.41&0.99\\
51916 &150.48936&2.688268&0.657&spec-$z$&$13\times2\times 60$&$15\times2\times120$&$ 8\times2\times120$&0.76&10.66&10.93&0.00\\
51982 &150.40459&2.780556&0.595&phot-$z$&$ 4\times2\times 60$&$ 5\times2\times120$&$ 4\times2\times120$&1.33&10.79&10.82&1.00
\enddata
\tablecomments{ 
Column (1): ID in the S-COSMOS MIPS 24 \um\ catalog. X-ray sources are indicated by stars.
Column (2): Right Ascension of the optical counterpart.
Column (3): Declination of the optical counterpart.
Column (4): Redshift.
Column (5): Redshift flag.
Columns (6)$-$(8): Exposure times for the 3 IRS modules; the total integration time equals the multiplication of the number of cycles, the number of dithering positions of each cycle, and the exposure time of an individual frame (in seconds). 
Column (9): MIPS 24 \um\ flux in mJy, these are before the color correction described in \S~\ref{irsobs}. 
Columns (10)-(11): Rest-frame luminosities in Log(\lsun) at 15 \um\ and
IRAC 8 \um; directly measured from the IRS spectra.
Column (12): Contribution of the AGN component to the integrated rest-frame
5.5 to 20 \um\ luminosity.
Sources outside of our redshift bin (\zbin) are included for completeness.
}
\end{deluxetable}

\begin{deluxetable}{cccccccc} 
\setlength{\tabcolsep}{.11cm}
\tablewidth{0pt}
\tablecaption{Star-forming 15/8\,\um\ Luminosity Functions \label{tab:SFLF}}
\tablehead{ 
\colhead{\Lfif} & \colhead{$N$(15\um)} & \colhead{$\phi$(15\um)} & \colhead{$\delta\phi$(15\um)} & 
\colhead{\Leig} & \colhead{$N$(8\um)} & \colhead{$\phi$(8\um)} & \colhead{$\delta\phi$(8\um)} 
\\
\colhead{(1)} & \colhead{(2)} & \colhead{(3)} & \colhead{(4)} & 
\colhead{(5)} & \colhead{(6)} & \colhead{(7)} & \colhead{(8)} 
}
\startdata
\hline
\multicolumn{8}{c}{MIPS Sample}   \nl
\hline
 9.68&1017&-2.39&-3.89& 9.94& 873&-2.33&-3.78\\
 9.82& 850&-2.50&-3.97&10.10& 938&-2.44&-3.93\\
 9.98& 675&-2.60&-4.02&10.24& 757&-2.55&-3.99\\
10.12& 468&-2.76&-4.10&10.40& 548&-2.69&-4.06\\
10.27& 283&-2.98&-4.21&10.55& 386&-2.85&-4.14\\
10.43& 173&-3.20&-4.31&10.69& 224&-3.08&-4.26\\
10.57&  69&-3.59&-4.51&10.85& 117&-3.37&-4.40\\
\hline
\multicolumn{8}{c}{IRS Sample} \nl
\hline
10.57&   9&-3.61&-4.01&10.85&   4&-3.33&-3.39\\
10.73&  14&-4.10&-4.67&10.99&  16&-3.85&-4.44\\
10.88&   4&-4.66&-4.96&11.15&   6&-4.47&-4.85\\
11.02&   1&-5.26&-5.26&11.30&   1&-5.26&-5.26
\enddata
\tablecomments{ 
Columns (1,5): Luminosity bin center.
Columns (2,6): Number of objects in the bin.
Columns (3,7): Comoving volume density in log(Mpc$^{-3}$ dex$^{-1}$).
Columns (4,8): Poisson error of the comoving volume density in log(Mpc$^{-3}$ dex$^{-1}$).
}
\end{deluxetable}

\begin{deluxetable}{cccccccc} 
\setlength{\tabcolsep}{.11cm}
\tablewidth{0pt}
\tablecaption{AGN 15/8\,\um\ Luminosity Functions \label{tab:AGNLF}}
\tablehead{ 
\colhead{\Lfif} & \colhead{$N$(15\um)} & \colhead{$\phi$(15\um)} & \colhead{$\delta\phi$(15\um)} & 
\colhead{\Leig} & \colhead{$N$(8\um)} & \colhead{$\phi$(8\um)} & \colhead{$\delta\phi$(8\um)} 
}
\startdata
\hline
10.65&   3&-4.18&-4.32&10.77&   4&-4.58&-4.88\\
10.80&   6&-4.44&-4.83&10.92&   3&-4.78&-5.02\\
10.95&   3&-4.78&-5.02&11.07&   3&-4.78&-5.02\\
11.10&   3&-4.78&-5.02&11.22&   2&-4.96&-5.11\\
11.25&   2&-4.96&-5.11&11.37&   1&-5.26&-5.26\\
11.40&   1&-5.26&-5.26&11.52&   1&-5.26&-5.26\\
11.70&   1&-5.26&-5.26&11.67&   1&-5.26&-5.26
\enddata
\end{deluxetable}

\begin{deluxetable}{lcccl} 
\tablecaption{Best-Fit Parameters of Star-forming Luminosity Functions \label{tab:fit}}
\tablehead{ \colhead{Parameter} & \colhead{15\,\um} & \colhead{IRAC 8\,\um} & \colhead{8$-$1000\,\um} & \colhead{Unit} }
\startdata
\hline
$L^{\star}$   &   $10.31\pm0.04$ &   $10.53\pm0.04$ &   $11.58\pm0.03$ & log(\lsun) \\
$\phi^{\star}$&   $-2.38\pm0.05$ &   $-2.42\pm0.06$ &   $-2.72\pm0.04$ & log(Mpc$^{-3}$ dex$^{-1}$) \\
$\alpha$      &   $-1.25\pm0.07$ &   $-1.32\pm0.07$ &   $-1.46\pm0.05$ & \nodata \\
$\sigma$      &          \nodata &        \nodata   &\phn$0.30\pm0.01$ & \nodata \\
$\rho$$^1$    &\phn$7.66\pm0.02$ &\phn$7.88\pm0.02$ &\phn$8.85\pm0.02$ & log(\lsun\ Mpc$^{-3})$ 
\enddata
\tablecomments{
$^1$ Total integrated luminosity from the analytic fit. }
\end{deluxetable}

\begin{deluxetable}{cccc} 
\setlength{\tabcolsep}{.11cm}
\tablewidth{0pt}
\tablecaption{8$-$1000\,\um\ Luminosity Function \label{tab:IRLF}}
\tablehead{ 
\colhead{\Lir} & \colhead{$N$} & \colhead{$\phi$} & \colhead{$\delta\phi$} 
}
\startdata
\hline
11.60&  10&-2.83&-3.21\\
11.75&  44&-3.08&-3.86\\
11.90&  46&-3.44&-4.23\\
12.05&  27&-3.78&-4.43\\
12.20&  11&-4.38&-4.90\\
12.35&   6&-4.66&-5.04\\
12.50&   1&-5.43&-5.43\\
12.80&   1&-5.43&-5.43
\enddata
\end{deluxetable}

\end{document}